\RequirePackage[columnwise]{lineno}
\documentclass[aps,prl,twocolumn,superscriptaddress,citeautoscript,longbibliography]{revtex4-2}
\usepackage{graphicx}  
\graphicspath{{figures/}} 
\usepackage{dcolumn}   
\usepackage{bm}        
\usepackage{braket}
\usepackage{exscale}                  
\usepackage[intlimits]{amsmath}       
\usepackage{amsfonts}
\usepackage{amssymb,amscd}
\usepackage{color}
\usepackage[colorlinks=true,linkcolor=blue,citecolor=blue,urlcolor=blue]{hyperref}
\usepackage{subfigure}
\usepackage[normalem]{ulem}
\usepackage{slashed}
\usepackage{leftidx}
\usepackage{xcolor,soul}
\usepackage{float}
\usepackage{easyReview}
\usepackage{multirow}

\soulregister\cite7
\soulregister\ref7
\definecolor{PowderBlue}{cmyk}{0.1137,0.0039,0,0.0627}
\sethlcolor{PowderBlue}
\AtBeginDocument{\let\oldcontentsline\contentsline}
\newcommand{\notoccontentsline}[4]{\oldcontentsline{}{}{}{}}
\newcommand{\droptocpage}{\addtocontents{toc}{\let\protect\contentsline\protect\notoccontentsline}}
\newcommand{\incltocpage}{\addtocontents{toc}{\let\protect\contentsline\protect\oldcontentsline}}

\begin{document}

%%%%%% title
\title{Functional building blocks for scalable multipartite entanglement in optical lattices}

%%%%%% authors
\author{Wei-Yong Zhang}
\thanks{These authors contributed equally to this work.}
\affiliation{Hefei National Research Center for Physical Sciences at the Microscale and School of Physics, University of Science and Technology of China, Hefei 230026, China}
\affiliation{CAS Center for Excellence in Quantum Information and Quantum Physics, University of Science and Technology of China, Hefei 230026, China}

\author{Ming-Gen He}
\thanks{These authors contributed equally to this work.}
\affiliation{Hefei National Research Center for Physical Sciences at the Microscale and School of Physics, University of Science and Technology of China, Hefei 230026, China}
\affiliation{CAS Center for Excellence in Quantum Information and Quantum Physics, University of Science and Technology of China, Hefei 230026, China}

\author{Hui Sun}
\thanks{These authors contributed equally to this work.}
\affiliation{Hefei National Research Center for Physical Sciences at the Microscale and School of Physics, University of Science and Technology of China, Hefei 230026, China}
\affiliation{CAS Center for Excellence in Quantum Information and Quantum Physics, University of Science and Technology of China, Hefei 230026, China}

\author{Yong-Guang Zheng}
\affiliation{Hefei National Research Center for Physical Sciences at the Microscale and School of Physics, University of Science and Technology of China, Hefei 230026, China}
\affiliation{CAS Center for Excellence in Quantum Information and Quantum Physics, University of Science and Technology of China, Hefei 230026, China}

\author{Ying Liu}
\affiliation{Hefei National Research Center for Physical Sciences at the Microscale and School of Physics, University of Science and Technology of China, Hefei 230026, China}
\affiliation{CAS Center for Excellence in Quantum Information and Quantum Physics, University of Science and Technology of China, Hefei 230026, China}

\author{An Luo}
\affiliation{Hefei National Research Center for Physical Sciences at the Microscale and School of Physics, University of Science and Technology of China, Hefei 230026, China}
\affiliation{CAS Center for Excellence in Quantum Information and Quantum Physics, University of Science and Technology of China, Hefei 230026, China}

\author{Han-Yi Wang}
\affiliation{Hefei National Research Center for Physical Sciences at the Microscale and School of Physics, University of Science and Technology of China, Hefei 230026, China}
\affiliation{CAS Center for Excellence in Quantum Information and Quantum Physics, University of Science and Technology of China, Hefei 230026, China}

\author{Zi-Hang Zhu}
\affiliation{Hefei National Research Center for Physical Sciences at the Microscale and School of Physics, University of Science and Technology of China, Hefei 230026, China}
\affiliation{CAS Center for Excellence in Quantum Information and Quantum Physics, University of Science and Technology of China, Hefei 230026, China}

\author{Pei-Yue Qiu}
\affiliation{Hefei National Research Center for Physical Sciences at the Microscale and School of Physics, University of Science and Technology of China, Hefei 230026, China}
\affiliation{CAS Center for Excellence in Quantum Information and Quantum Physics, University of Science and Technology of China, Hefei 230026, China}

\author{Ying-Chao Shen}
\affiliation{Hefei National Research Center for Physical Sciences at the Microscale and School of Physics, University of Science and Technology of China, Hefei 230026, China}
\affiliation{CAS Center for Excellence in Quantum Information and Quantum Physics, University of Science and Technology of China, Hefei 230026, China}

\author{Xuan-Kai Wang}
\affiliation{Hefei National Research Center for Physical Sciences at the Microscale and School of Physics, University of Science and Technology of China, Hefei 230026, China}
\affiliation{CAS Center for Excellence in Quantum Information and Quantum Physics, University of Science and Technology of China, Hefei 230026, China}

\author{Wan Lin}
\affiliation{Hefei National Research Center for Physical Sciences at the Microscale and School of Physics, University of Science and Technology of China, Hefei 230026, China}
\affiliation{CAS Center for Excellence in Quantum Information and Quantum Physics, University of Science and Technology of China, Hefei 230026, China}

\author{Song-Tao Yu}
\affiliation{Hefei National Research Center for Physical Sciences at the Microscale and School of Physics, University of Science and Technology of China, Hefei 230026, China}
\affiliation{CAS Center for Excellence in Quantum Information and Quantum Physics, University of Science and Technology of China, Hefei 230026, China}

\author{Bin-Chen Li}
\affiliation{Hefei National Research Center for Physical Sciences at the Microscale and School of Physics, University of Science and Technology of China, Hefei 230026, China}
\affiliation{CAS Center for Excellence in Quantum Information and Quantum Physics, University of Science and Technology of China, Hefei 230026, China}

\author{Bo Xiao}
\affiliation{Hefei National Research Center for Physical Sciences at the Microscale and School of Physics, University of Science and Technology of China, Hefei 230026, China}
\affiliation{CAS Center for Excellence in Quantum Information and Quantum Physics, University of Science and Technology of China, Hefei 230026, China}

\author{Meng-Da Li}
\affiliation{Hefei National Research Center for Physical Sciences at the Microscale and School of Physics, University of Science and Technology of China, Hefei 230026, China}
\affiliation{CAS Center for Excellence in Quantum Information and Quantum Physics, University of Science and Technology of China, Hefei 230026, China}

\author{Yu-Meng Yang}
\affiliation{Hefei National Research Center for Physical Sciences at the Microscale and School of Physics, University of Science and Technology of China, Hefei 230026, China}
\affiliation{CAS Center for Excellence in Quantum Information and Quantum Physics, University of Science and Technology of China, Hefei 230026, China}

\author{Xiao Jiang}
\affiliation{Hefei National Research Center for Physical Sciences at the Microscale and School of Physics, University of Science and Technology of China, Hefei 230026, China}
\affiliation{CAS Center for Excellence in Quantum Information and Quantum Physics, University of Science and Technology of China, Hefei 230026, China}

\author{Han-Ning Dai}
\affiliation{Hefei National Research Center for Physical Sciences at the Microscale and School of Physics, University of Science and Technology of China, Hefei 230026, China}
\affiliation{CAS Center for Excellence in Quantum Information and Quantum Physics, University of Science and Technology of China, Hefei 230026, China}

\author{You Zhou}
\affiliation{Hefei National Research Center for Physical Sciences at the Microscale and School of Physics, University of Science and Technology of China, Hefei 230026, China}
\affiliation{Key Laboratory for Information Science of Electromagnetic Waves (Ministry of Education), Fudan University, Shanghai 200433, China}

\author{Xiongfeng Ma}
\affiliation{Center for Quantum Information, Institute for Interdisciplinary Information Sciences, Tsinghua University, Beijing 100084, China}

\author{Zhen-Sheng~Yuan}
% \email[e-mail:]{yuanzs@ustc.edu.cn}
\affiliation{Hefei National Research Center for Physical Sciences at the Microscale and School of Physics, University of Science and Technology of China, Hefei 230026, China}
\affiliation{CAS Center for Excellence in Quantum Information and Quantum Physics, University of Science and Technology of China, Hefei 230026, China}
\affiliation{Hefei National Laboratory, University of Science and Technology of China, Hefei 230088, China}

\author{Jian-Wei~Pan}
% \email[e-mail:]{pan@ustc.edu.cn}
\affiliation{Hefei National Research Center for Physical Sciences at the Microscale and School of Physics, University of Science and Technology of China, Hefei 230026, China}
\affiliation{CAS Center for Excellence in Quantum Information and Quantum Physics, University of Science and Technology of China, Hefei 230026, China}
\affiliation{Hefei National Laboratory, University of Science and Technology of China, Hefei 230088, China}

%% Abstract 
\begin{abstract}
    Featuring excellent coherence and operated parallelly, ultracold atoms in optical lattices form a competitive candidate for quantum computation. For this, a massive number of parallel entangled atom pairs have been realized in superlattices. However, the more formidable challenge is to scale-up and detect multipartite entanglement due to the lack of manipulations over local atomic spins in retro-reflected bichromatic superlattices. Here we developed a new architecture based on a cross-angle spin-dependent superlattice for implementing layers of quantum gates over moderately-separated atoms incorporated with a quantum gas microscope for single-atom manipulation. We created and verified functional building blocks for scalable multipartite entanglement by connecting Bell pairs to one-dimensional 10-atom chains and two-dimensional plaquettes of $2\times4$ atoms. This offers a new platform towards scalable quantum computation and simulation.
\end{abstract}

\date{\today}
\maketitle
\droptocpage

\section{Introduction}

Entanglement is the key resource for applications in quantum information science \cite{Nielsen:2010qc,Pezze:2018qm,Raussendorf:2001ao}. To achieve the long-term pursuit to realize scalable multipartite entanglement, remarkable advances have been made in different quantum platforms including photons, superconducting circuits, trapped ions, and neutral atoms in tweezer arrays \cite{Wang:2018qe,Gong:2019gq,Song:2019go,wang:201816,mooney:2021ga,friis:2018oo,Omran:2019ga,Bluvstein:2022aq,Graham:2022mq}. An alternative system, ultracold atoms in optical lattices, can promisingly enable a programmable approach to generate scalable two-dimensional (2D) entanglement thanks to a large number of atomic qubits generated through a quantum phase transition and highly parallel spin operations \cite{Jaksch:1998cb,Greiner:2002qp,Duan:2003cs,Trotzky:2008tr,Dai:2016ga}. Recently, progress has been achieved in manipulating atomic qubits in optical lattices \cite{Chiu:2018qs,Yang:2020ca,Weitenberg:2011ss,Bakr:2009aq,Sherson:2010sa,Gross:2021qg}. This advances support a prospective research route to scale up multipartite entanglement with quantum gates provided by controllable bichromatic superlattices \cite{Vaucher:2008co}, where the landmarks include generating isolated entangled Bell pairs in parallel, entangling Bell pairs into one-dimensional (1D) chains, and further realizing 2D entangled plaquettes. As a first step, the parallel preparation of isolated Bell pairs has been demonstrated with quantum gates in tunable superlattices \cite{Dai:2016ga,Yang:2020ca}.

The next critical step is to realize the cascade of entangling gates applied on configurable qubit arrays along the two spatial dimensions, leading to the programmable generation of 2D multipartite spin entanglement. This approach requires the combination of all the fundamental operations including local spin manipulation, parallel controllable entangling gates and single-site-resolved detection of spin states. However, the small atom separation in conventional retroreflected bichromatic superlattices \cite{Yang:2017sd} is beyond the current imaging resolution in systems of neutral alkali atoms \cite{Bakr:2009aq,Sherson:2010sa}. In addition, the accumulating noise of the quantum gates and the decoherence of the qubits can degrade the final multipartite entanglement, limiting the effectiveness of quantum circuits in optical lattices.

\begin{figure*}[htb]
    \centering
    \includegraphics[width=170mm]{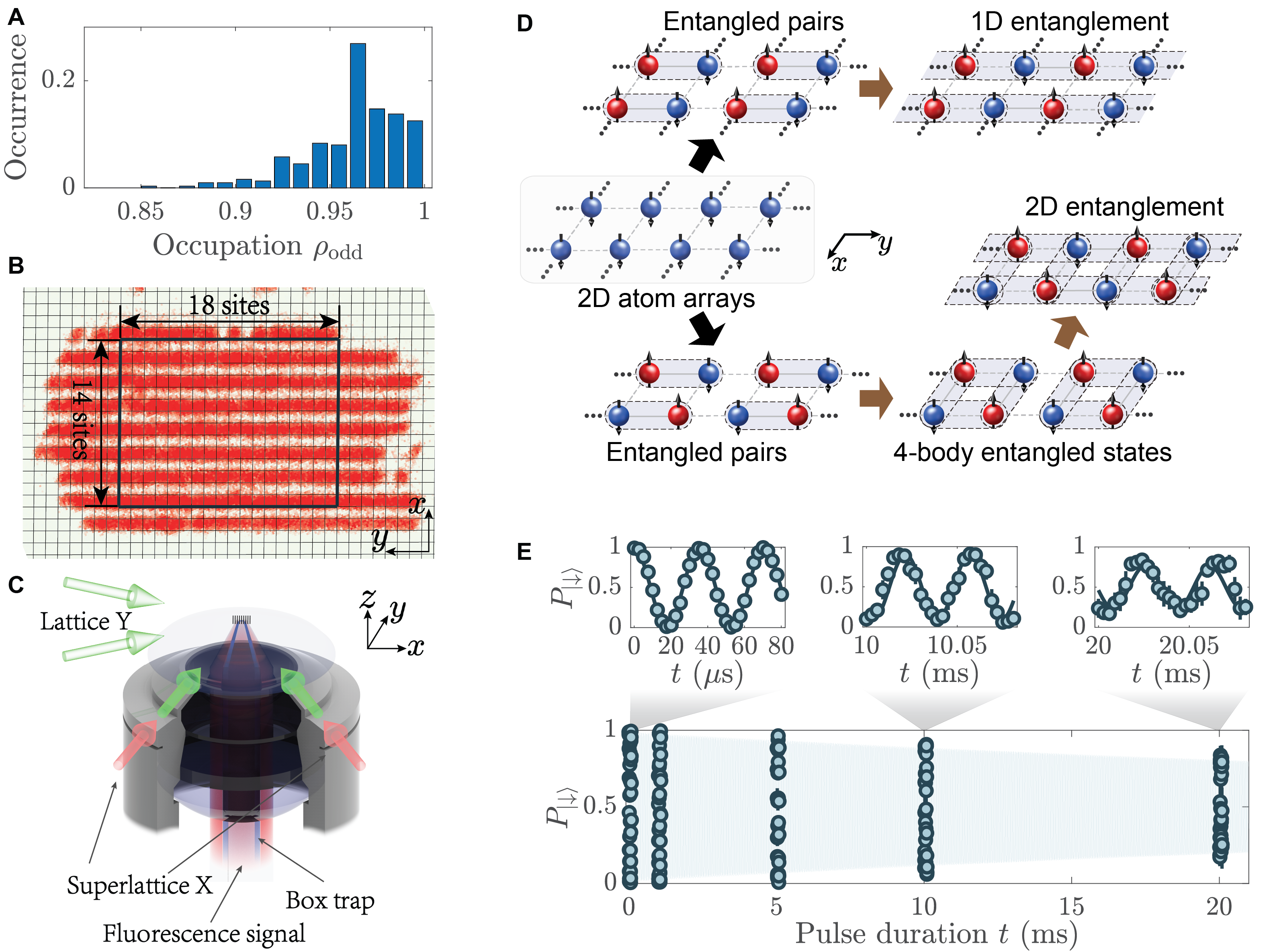}
    \caption{\textbf{Experimental setup and sketch for entanglement generation}. 
    (\textbf{A}) Histogram of the average occupation in the region of interest (ROI) as marked in (B) after cooling atoms in the superlattice. (\textbf{B}) An exemplary image of the atom array and the marked ROI containing $18 \times 14$ lattice sites. (\textbf{C}) Simplified experimental setup. Ultracold $^{87}$Rb atoms are loaded into a 2D staggered superlattice (see Methods). A digital micro-mirror device (DMD) is used to create a box trap and meanwhile to compensate the inhomogeneity of the harmonic trap. The fluorescence of atoms is collected by the objective along the $z$-direction. (\textbf{D}) A cartoon schematic of preparing multipartite entanglement in optical lattices. (\textbf{E}) Measured occupation of the $\ket{\downarrow}$ state in the driven Rabi oscillation via the Stern-Gerlach-type approach (see Methods), indicating a exponential decay with a time constant of $\tau_{\mathrm{Rabi}} = 42.9 \pm 7.0 ~\mathrm{ms}$.
    }
    \label{figure1:stagC}
\end{figure*}

Here we demonstrate the functional building blocks for generate and probe scalable multipartite atomic entanglement in optical lattices. This is realized by developing a cross-angle spin-dependent optical superlattice for trapping moderately separated atoms incoparated with a quantum gas microscope for single-atom manipulation. Parallel and local spin controllability are developed by combining the spin-dependent superlattice and versatile atom addressing techniques in the single-site precision using digital micromirror devices (DMDs) \cite{Zupancic:2016up}. Actively reducing the noises of magnetic fields and lasers, long-lived Bell pairs with a lifetime of $2.20 \pm 0.13$ s are prepared using parallel high-fidelity entangling gates based on the superexchange effect in double wells. We then demonstrate the programmable generation of 1D and 2D multipartite entangled states by entangling a ten-atom chain and a plaquette of $2 \times 4$ atoms. The full bipartite non-separability of these states is experimentally verified with site- and spin-resolved detection methods. 

\begin{figure*}[ht!]
    \centering     %
    \includegraphics[width=175mm]{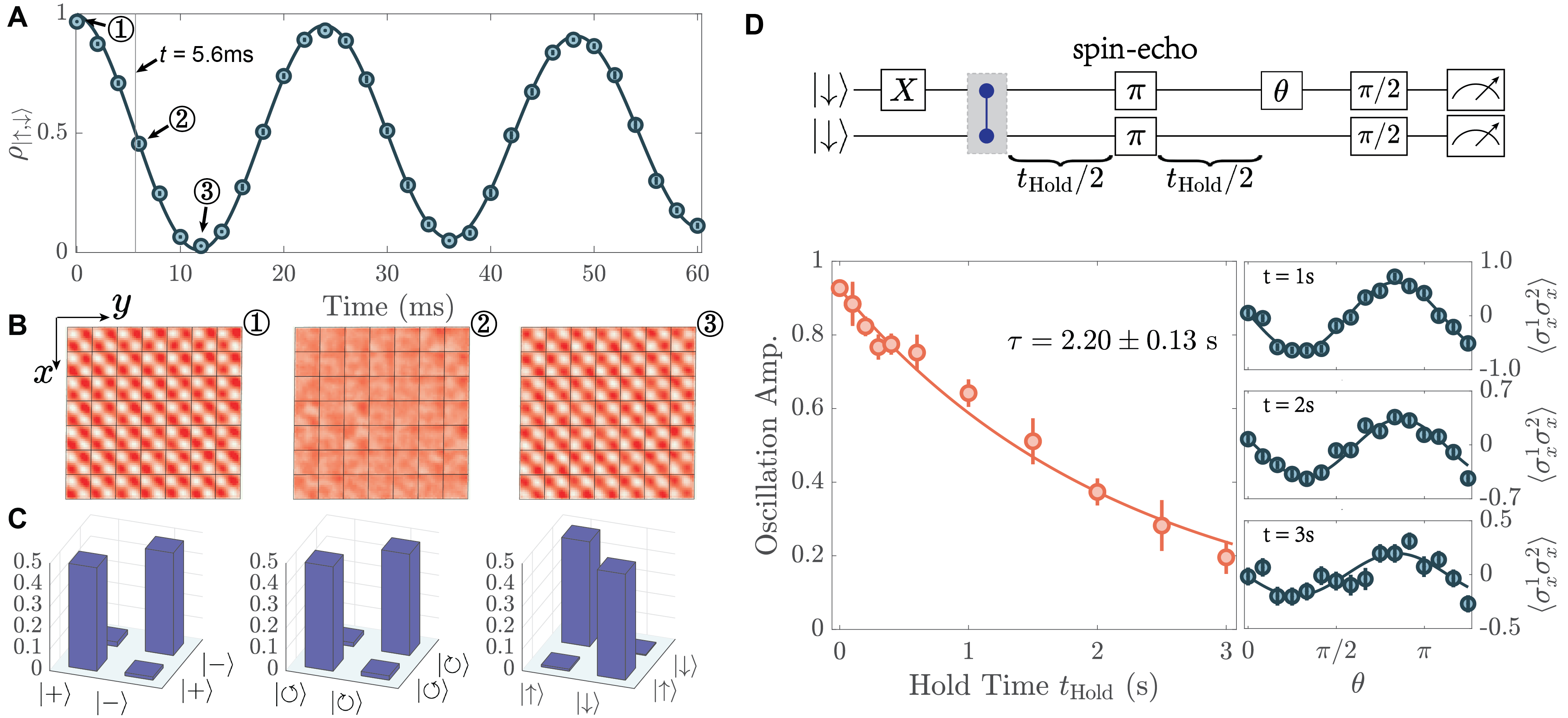}
    \caption{\textbf{Superexchange dynamics and Bell states}.
    (\textbf{A}) The averaged occupancy (circles) of the $\ket{\uparrow,\downarrow}$ state, showing the superexchange dynamics in isolated double-wells within the ROI. The fitted curve (solid line) with a damped sinusoidal function gives the averaged exchange strength $J_{\mathrm{ex}} = 20.5 \pm 0.1$ Hz. (\textbf{B}) The averaged spin configurations measured in the 49 plaquettes during the spin superexchange process in double-wells. (\textbf{C}) Measured populations under $\ket{+/-}$, $\ket{\circlearrowleft / \circlearrowright}$ and $\ket{\uparrow / \downarrow}$ basis for the prepared Bell states averaged within the ROI. Gives $P_{+,+} = 0.484 \pm 0.007$, $P_{+,-} = 0.021 \pm 0.003$, $P_{-,+} = 0.015 \pm 0.002$, $P_{-,-} = 0.480 \pm 0.007$, $P_{\circlearrowleft,\circlearrowleft} = 0.488 \pm 0.010$, $P_{\circlearrowleft,\circlearrowright} = 0.018 \pm 0.002$, $P_{\circlearrowright,\circlearrowleft} = 0.020 \pm 0.003$, $P_{\circlearrowright,\circlearrowright} = 0.474 \pm 0.008$, $P_{\uparrow,\uparrow} = 0.010 \pm 0.001$, $P_{\uparrow,\downarrow} = 0.490 \pm 0.009$, $P_{\downarrow,\uparrow} = 0.496 \pm 0.009$, $P_{\downarrow,\downarrow} = 0.004 \pm 0.001$. (\textbf{D}) The extracted parity oscillation contrast of the prepared Bell states after different holding times with a spin-echo $\pi$ pulse, fitted as an  exponential decay with a time constant $\tau = 2.20 \pm 0.13 ~\mathrm{s}$. Error bars denote the s.e.m.
    }
    \label{figure2:xSE}
\end{figure*}

\section{Preparing a defect-free 2D qubit array}

In the experiment, ultracold $^{87}\mathrm{Rb}$ atoms are loaded into a 2D far-detuned optical superlattice generated by equal-arm interferometers, which are constructed by 532-nm/1064-nm lasers for short/long lattices \cite{Li:2021hp}. The lattice spacing is 630 nm resulting from the cross angle of 50 degree between the short lattices, well resolved with our imaging resolution. This equal-arm configuration, resistant to external perturbations, leading to the long-term stability of the absolute location of the lattice sites. By tuning the spin dependency and relative phases of the superlattice, the experimental setup is further equipped with the parallel and local manipulation over atomic spin states and the single-site-resolved detection (see Methods), which enable parallel assembly of large-scale defect-free qubits with the recently demonstrated cooling method \cite{Yang:2020ca}. In addition, we apply a repulsive optical potential projected by a digital micromirror device (DMD) to compensate the harmonic confinement originating from the Gaussian envelope of lattice beams. Thus, with an average site occupation of $\sim 0.75$ during the cooling process, a near-perfect unit-filling atom array is obtained in every second rows of short lattices after removing the superfluid reservoirs.

Fig.~\ref{figure1:stagC}B shows an exemplary distribution of the atomic parity in one shot and a marked region of interest (ROI) containing $18 \times 14$ lattice sites. Considering the detection loss during imaging, the overall fidelity for preparing a unity filling state across 200 lattice sites is 99.2(2)\% (the value in parentheses represents s.e.m, see Methods), obtained from around 400 experimental repetitions. The hyperfine ground states $\ket{\downarrow} = \ket{F=1,m_{\mathrm{F}}=-1}$ and $\ket{\uparrow} = \ket{F=2,m_{\mathrm{F}}=-2}$ are used to encode the qubit states. The atoms possess excellent single-qubit coherence, verified by driven Rabi oscillations accompanied by a exponential decay with a time constant of $\tau_{\mathrm{Rabi}} = 42.9 \pm 7.0 ~\mathrm{ms}$ in Fig.~\ref{figure1:stagC}E. 

\begin{figure*}[htb]
    \centering     %
    \includegraphics[width=175mm]{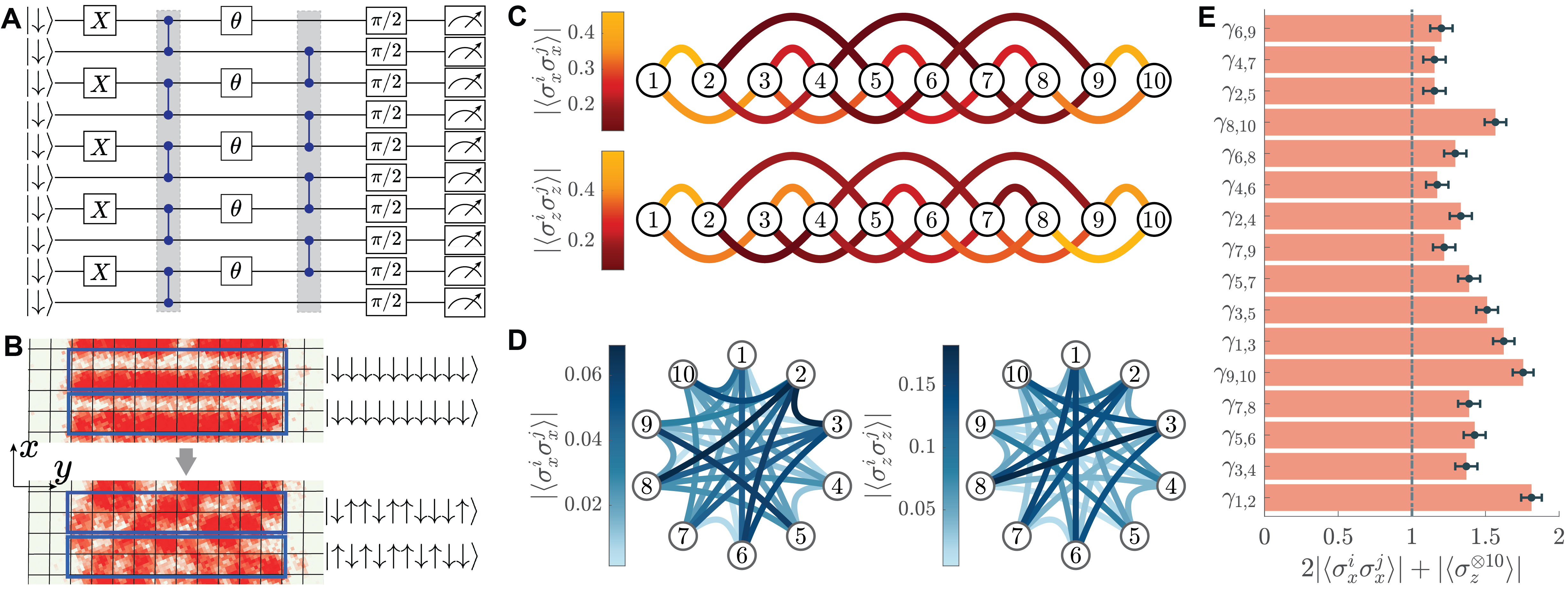}
    \caption{\textbf{Ten-body fully-entangled state}.
    (\textbf{A}) Quantum circuit representation for preparing and measuring ten-body fully entangled state. (\textbf{B}) Exemplary images of the measured spin configuration before and after the entanglement preparation under the $\sigma_z$ basis, respectively. (\textbf{C} and \textbf{D}) The measured two-particle spin correlations of the ten-atom fully entangled state under the $\sigma_z$ and $\sigma_x$ basis. (\textbf{E}) The extracted results of observables $\gamma_{i,j}$. The dashed dot line represents the threshold for the verification of the fully entangled state. Since $\gamma_{1,2} = 1.81 \pm 0.09$, $\gamma_{3,4} = 1.37 \pm 0.10$, $\gamma_{5,6} = 1.43 \pm 0.10$, $\gamma_{7,8} = 1.39 \pm 0.10$, and $\gamma_{9,10} = 1.76 \pm 0.09$, we conclude that atoms 1 and 2, 3 and 4, 5 and 6, 7 and 8, 9 and 10 are non-separable, respectively. Moreover, since $\gamma_{1,3} = 1.62 \pm 0.10$, $\gamma_{3,5} = 1.51 \pm 0.10$, $\gamma_{5,7} = 1.39 \pm 0.10$, and $\gamma_{8,10} = 1.57 \pm 0.10$, we deduce that the 10 atoms should be contained in $M$ or $\bar{M}$ (see Methods). Error bars denote the s.e.m.
    }
    \label{figure3:10body}
\end{figure*}

\section{Site- and spin-resolved state photography and entangled pairs}

With the defect-free qubit array, the entangling $\sqrt{\rm{SWAP}}^{\dagger}$ gate based on the spin superexchange interaction is investigated in a site-resolved way within the ROI containing $14 \times 14$ lattice sites. The experiment starts by first preparing the two $^{87}\mathrm{Rb}$ atoms in each isolated double-well in a N\'eel-type antiferromagnetic order $\ket{\uparrow,\downarrow}$ (the comma separating the left and right occupations). With the auxiliary unoccupied lattice sites alternately arranged along the $x$-direction, we can apply a Stern-Gerlach-type single-spin-resolved detection (see Methods) to directly visualize the dynamic evolution of these four spin states, $\{ \ket{\uparrow,\downarrow}, \ket{\downarrow,\uparrow}, \ket{\uparrow \downarrow,0}, \ket{0,\uparrow \downarrow} \}$ \cite{Boll:2016sa}. The fidelity of this detection method is around 99.5\% (see more details in Tab.~\ref{table:errorbudget}) owing to the remarkable stability of our optical superlattice.

Fig.~\ref{figure2:xSE}A shows the measured time-resolved average occupancy of the $\ket{\uparrow,\downarrow}$ state over the 49 double wells (DWs), which reveals the expected sinusoidal oscillation at a frequency of $J_{\mathrm{ex}}/h = 20.5 \pm 0.1 ~\mathrm{Hz}$. The residual inhomogeneity of trapping potential induces a slight deviation in the synchronization in different double wells (as shown in Fig.~\ref{figureS8:JexMap}) during the evolution, which, together with the superlattice phase fluctuations and residual spin-dependent effects of lattices, shortens the lifetime of superexchange dynamics. 

We then prepare two-atom entangled Bell states in parallel using entangling gates $\sqrt{\mathrm{SWAP}}^{\dagger}$. This is realized by halting the superexchange dynamics at a fixed evolution time of $t = 5.6~\mathrm{ms}$, followed by a phase rotation  \cite{Trotzky:2010ca}. From the measured two-atom spin correlations shown in Fig.\ref{figure2:xSE}C, we characterize the average fidelity of Bell states in the ROI as $\mathcal{F} = 0.956 \pm 0.005$. The coherence time of the Bell states is obtained by the Ramsey interference measurement, implemented with the singlet-triplet oscillation (STO) after variable holding time. As shown in Fig.~\ref{figure2:xSE}D, the extracted oscillation amplitudes exhibit the exponential decay with a time constant of $\tau = 2.20 \pm 0.13$ s \cite{Yang:2019ul,Xiao:2020gt,Li:2021ah}. This long coherence time guarantees scaling up the entanglement of atoms. Being equivalent to the CNOT gate up to several single-qubit rotations, our $\sqrt{\mathrm{SWAP}}^{\dagger}$ gate can serve as an elementary two-qubit gate for universal quantum computation \cite{Barenco:1995eg,Loss:1998qc,Nielsen:2010qc}.

\begin{figure*}[htb]
    \centering     %
    \includegraphics[width=175mm]{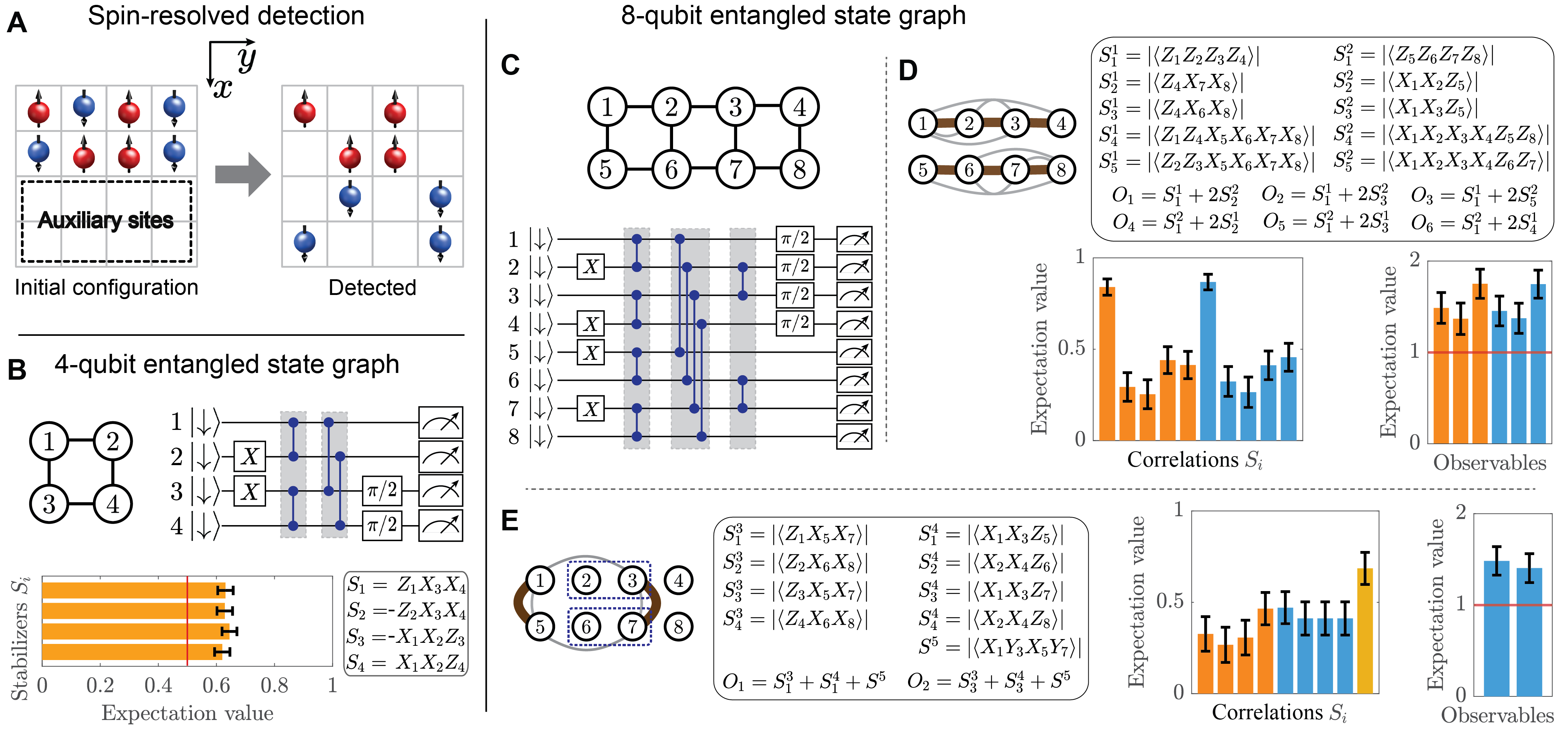}
    \caption{\textbf{2D four-body and eight-body entangled states}. 
    (\textbf{A}) The schematic sketch of fully spin-resolved detection in a 2D configuration, which is achieved via state-dependent atom transport in optical lattices. (\textbf{B}) Top panel: The spatial structure of a 2D, four-body entangled state and the corresponding quantum circuit representation of state preparation and measurement. Bottom panel: Measured spin correlations and observables for demonstrating four-body genuine entanglement. (\textbf{C}) Spatial structure of a 2D, eight-body entangled plaquette and the corresponding quantum circuit representation of state preparation and measurement. (\textbf{D}) Measured spin correlations for the first two measurement settings and corresponding observables. These measurements certify the full entanglement inside two individual chains, shown with solid brown marks. (\textbf{E}) Measured spin correlations for the last three measurement settings and corresponding observables. These measurements certify the full entanglement between the two chains, shown with solid brown marks. All the results presented are not corrected for detection errors. The red solid lines in histograms represent the threshold for the bipartite non-separability. Error bars denote the s.e.m.
    }
    \label{figure4:2DEntangledS}
\end{figure*}

\section{Connecting entangled pairs to 1D and 2D multipartite entanglement}

The long coherence time of entangled pairs and the engineered superlattice potential together form the basis to cascade quantum gates across 1D and 2D qubit arrays. Besides, the phase stability of the lattices are crucial for addressing a desired pattern of single atoms.

We first demonstrate the generation of 1D entanglement by cascading two parallel layers of entangling gates. The initial state is prepared as a chain of ten qubits with the  N\'eel-order magnetization, where all other atoms in the array are removed by applying the site-resolved atom addressing techniques. Shown in Fig.~\ref{figure3:10body}A, after preparing isolated two-atom entangled Bell states, we change the relative phase $\theta_{y}$ of the superlattice to form new atom pairs, and then apply another parallel layer of $\sqrt{\mathrm{SWAP}}^{\dagger}$ gates. Since verifying the ten-qubit genuine entanglement requires a rather complicated measurement procedure, we develop an efficient entanglement criterion on the basis of the scheme \cite{Zhou:2022qg}, which can verify the full bipartite non-separability of the entangled chain (see Methods). According to the new criterion, we only need to measure the final spin configuration under two settings of bases, $\sigma_x^{\otimes 10}$ and $\sigma_z^{\otimes 10}$, given by whether to apply a global $\pi/2$ rotation pulse before the projection measurement. The measurement results of two-particle spin correlations are shown in Fig.~\ref{figure3:10body}, C and D. All the observables $\gamma_{i,j} =  2 \left| \left\langle \sigma_x^{i} \sigma_x^{j} \right\rangle \right| + \left| \left\langle \otimes_{k=1}^{10} \sigma_z^{k} \right\rangle \right|$, shown in Fig.~\ref{figure3:10body}E, surpass the threshold, verifying that the ten-atom chain cannot be divided into any two separated partitions. Therefore, we realize a ten-atom fully entangled state with the quantum circuit involving two layers of entangling gates.

Next, we demonstrate the 2D gate-based quantum circuit of generating and measuring fully entangled states of four atoms in a $2\times2$ plaquette and eight atoms in a $2\times4$ plaquette, respectively.

Since the Stern-Gerlach techniques are no longer applicable for measuring 2D entanglement, we develop a new detection approach to resolve the atomic spins on each site using state-dependent atom transport. As illustrated in Fig.~\ref{figure4:2DEntangledS}A, we first shine a state-dependent addressing beam shaped by a DMD to pin the $\ket{\uparrow}$  atoms, and then change the phase of the long lattice to transport the $\ket{\downarrow}$ atoms to the originally unoccupied auxiliary sites before performing fluorescence imaging. Atoms in the two spin states can be distinguished simultaneously by their final positions. As in Tab.~\ref{table:errorbudget}, the average fidelity of this detection technique is around 98.5\%, where the inefficiency is attributed to the influence of cross-talking between different spin states and accidental atom hoppings during the detection.

Fig.~\ref{figure4:2DEntangledS}B shows the preparation of four-body entanglement in isolated $2 \times 2$ plaquettes. We begin with a 2D N\'eel-order magnetization. After the first parallel layer of $\sqrt{\mathrm{SWAP}}^{\dagger}$ gates generate Bell pairs along the $y$-direction, we realize the four-body entangled state by applying a second layer of $\sqrt{\mathrm{SWAP}}^{\dagger}$ gates along the $x$-direction. The target state is a stabilizer state. The generators of its stabilizer group are $\sigma_z^1 \sigma_z^2$, $\sigma_z^3 \sigma_z^4$, $\sigma_z^1 \sigma_x^3 \sigma_x^4$ and $-\sigma_x^1 \sigma_x^2 \sigma_z^3$. To characterize this state, the spin correlations extracted from the two measurement settings, $\sigma_x^1 \sigma_x^2 \sigma_z^3 \sigma_z^4$ and $\sigma_z^1 \sigma_z^2 \sigma_x^3 \sigma_x^4$, are obtained by performing projective measurements after a local $\pi/2$ pulse is applied to two bottom sites or the two top sites (see Methods). The measured expectation values, as shown in Fig.~\ref{figure4:2DEntangledS}B, satisfy the inequality Eq.~\ref{equation:S11}, thus verify the four-body genuine entanglement.

Furthermore, we realize the 2D eight-body entanglement in the $2\times4$ plaquette by applying one more layer of entangling gates to connect neighboring $2 \times 2$ plaquettes, as shown in Fig.~\ref{figure4:2DEntangledS}C. After entangling atoms in isolated $2 \times 2$ plaquette as above, we switch the relative phase $\theta_y$ of the superlattice along the $y$-direction and further apply another layer of parallel $\sqrt{\mathrm{SWAP}}^{\dagger}$ gates to entangle these adjacent plaquettes. With the following steps, we characterize this state by verifying the full bipartite non-separability. First, we perform the same measurements as above for demonstrating four-body entangled state in isolated plaquettes. Shown in Fig.~\ref{figure4:2DEntangledS}D, the yielded observables surpass the threshold, certifying the full entanglement property inside the two separate chains along the $y$-direction. Then, we demonstrate non-separability between these two chains. To build proper measurement bases, we apply an additional layer of $\sqrt{\mathrm{SWAP}}^{\dagger}$ gates along the $y$-direction, which is a LOCC (local operations and classical communication) inside each chain and cannot enhance entanglement between the two chains. Thus, after these auxiliary gates, we build an observable verifying the interchain entanglement with three measurements. After two measurements the same as above, we further measure the spin correlation $\sigma_x^1 \sigma_y^3 \sigma_x^5 \sigma_y^7$ by applying a $\pi/2$ pulse with site-dependent phases (see Methods). Shown in Fig.~\ref{figure4:2DEntangledS}E, all necessary observables surpass the relevant classical thresholds, leading to the verfication of 2D eight-body full entanglement.

In Fig.~\ref{figure3:10body}C, we can see an approximate reflection symmetry in the measured two-body spin correlations, revealing the symmetry of the target states. The slight spatial-dependent deviations, as in Fig.~\ref{figure3:10body}, C and D, result from the remaining inhomogeneity of the overall trapping potentials and the residual magnetic gradient which contribute to the infidelity of the generated multipartite entanglement, as observed in Fig.~\ref{figure3:10body}E and Fig.~\ref{figure4:2DEntangledS}, B, D and E. These imperfections can be overcome by optimizing the projected compensation pattern from the DMD and purifying the lattice laser polarization. Besides, the efficiency of 2D spin-resolved detection can be enhanced by a more reliable state-dependent atom transport through upgrading the addressing technique. Such improvements may allow deep quantum circuits to entangle over hundreds of neutral atoms in a 2D configuration.

\section{Discussion and conclusion}

Our experiments demonstrate the essential ingredients in the roadmap for generating multipartite entangled states with two-qubit gates, from preparing isolated Bell pairs to realizing 1D entangled chains and 2D entangled plaquettes. High-fidelity quantum circuits are implemented by combining a quantum gas microscope with a cross-angle spin-dependent optical superlattice. Our experiment proves the capability of generating and detecting scalable entanglement in optical lattices. For implementing the measurement-based quantum computation (MBQC) \cite{Briegel:2009mb}, our two-qubit entangling gate can be employed to generate the cluster states with additional single-qubit rotation gates \cite{Vaucher:2008co,Tanamoto:2009eq}, which can be achieved after we further stabilize the lattice phases and smoothen the profile of the projected addressing beam. Besides, by integrating versatile controllable tight-focused optical traps \cite{zhang:2006mo,Weitenberg:2011qc}, we can also perform local measurements on individual atoms along different axes, satisfying the essential request of MBQC. More generally, our platform can offer new opportunities for quantum simulation of intriguing physics in lattice gauge theories \cite{Yang:2020oo,Zhou:2022td,Yao:2022qm,Cheng:2022tc,Halimeh:2022tt} and exotic quantum phases in the quantum magnetism realm \cite{Bohrdt:2021eo}. The capability on realizing low-entropy atom arrays together with the high-precision manipulation of single atoms may open the avenue to demonstrating practical quantum advantage \cite{Daley:2022pq}. 

%%%%%%%%% Acknowledgements
\textbf{Acknowledgement} 
We thank Otfried G\"{u}hne for discussions on entanglement verification, Juan Yin for suggestions on the thermal lensing effect of optics, and Qian Xie, Zhao-Yu Zhou and Guo-Xian Su for their early contribution to building the experimental setup. This work was supported by the NNSFC grant 12125409, the Innovation Program for Quantum Science and Technology 2021ZD0302000, and the Anhui Initiative in Quantum Information Technologies.

% \textbf{Competing interests:} 
%     The authors declare no competing interests.

% \textbf{Data availability:} 
%     Data used in this work is available on reasonable request.

% \textbf{Code availability:} 
%     Code used in this work is available on reasonable request.
    
% reference list
\bibliography{multi_ent}    % using bib formula for main body

\providecommand{\noopsort}[1]{}\providecommand{\singleletter}[1]{#1}%
\begin{thebibliography}{10}

\bibitem{Nielsen:2010qc}
M.~A. Nielsen, I.~L. Chuang, {\it Quantum Computation and Quantum Information:
  10th Anniversary Edition\/} (Cambridge University Press, 2010).

\bibitem{Pezze:2018qm}
L.~Pezze, A.~Smerzi, M.~K. Oberthaler, R.~Schmied, P.~Treutlein, {\it Reviews
  of Modern Physics\/} {\bf 90}, 035005 (2018).

\bibitem{Raussendorf:2001ao}
R.~Raussendorf, H.~J. Briegel, {\it Physical Review Letters\/} {\bf 86}, 5188
  (2001).

\bibitem{Wang:2018qe}
X.-L. Wang, {\it et~al.\/}, {\it Physical Review Letters\/} {\bf 120}, 260502
  (2018).

\bibitem{Gong:2019gq}
M.~Gong, {\it et~al.\/}, {\it Physical review letters\/} {\bf 122}, 110501
  (2019).

\bibitem{Song:2019go}
C.~Song, {\it et~al.\/}, {\it Science\/} {\bf 365}, 574 (2019).

\bibitem{wang:201816}
Y.~Wang, Y.~Li, Z.-Q. Yin, B.~Zeng, {\it npj Quantum Information\/} {\bf 4}, 46
  (2018).

\bibitem{mooney:2021ga}
G.~J. Mooney, G.~A. White, C.~D. Hill, L.~C. Hollenberg, {\it Journal of
  Physics Communications\/} {\bf 5}, 095004 (2021).

\bibitem{friis:2018oo}
N.~Friis, {\it et~al.\/}, {\it Physical Review X\/} {\bf 8}, 021012 (2018).

\bibitem{Omran:2019ga}
A.~Omran, {\it et~al.\/}, {\it Science\/} {\bf 365}, 570 (2019).

\bibitem{Bluvstein:2022aq}
D.~Bluvstein, {\it et~al.\/}, {\it Nature\/} {\bf 604}, 451 (2022).

\bibitem{Graham:2022mq}
T.~Graham, {\it et~al.\/}, {\it Nature\/} {\bf 604}, 457 (2022).

\bibitem{Jaksch:1998cb}
D.~Jaksch, C.~Bruder, J.~I. Cirac, C.~W. Gardiner, P.~Zoller, {\it Physical
  Review Letters\/} {\bf 81}, 3108 (1998).

\bibitem{Greiner:2002qp}
M.~Greiner, O.~Mandel, T.~Esslinger, T.~W. H{\"a}nsch, I.~Bloch, {\it nature\/}
  {\bf 415}, 39 (2002).

\bibitem{Duan:2003cs}
L.-M. Duan, E.~Demler, M.~D. Lukin, {\it Physical review letters\/} {\bf 91},
  090402 (2003).

\bibitem{Trotzky:2008tr}
S.~Trotzky, {\it et~al.\/}, {\it Science\/} {\bf 319}, 295 (2008).

\bibitem{Dai:2016ga}
H.-N. Dai, {\it et~al.\/}, {\it Nature Physics\/} {\bf 12}, 783 (2016).

\bibitem{Chiu:2018qs}
C.~S. Chiu, G.~Ji, A.~Mazurenko, D.~Greif, M.~Greiner, {\it Physical review
  letters\/} {\bf 120}, 243201 (2018).

\bibitem{Yang:2020ca}
B.~Yang, {\it et~al.\/}, {\it Science\/} {\bf 369}, 550 (2020).

\bibitem{Weitenberg:2011ss}
C.~Weitenberg, {\it et~al.\/}, {\it Nature\/} {\bf 471}, 319 (2011).

\bibitem{Bakr:2009aq}
W.~S. Bakr, J.~I. Gillen, A.~Peng, S.~F{\"o}lling, M.~Greiner, {\it Nature\/}
  {\bf 462}, 74 (2009).

\bibitem{Sherson:2010sa}
J.~F. Sherson, {\it et~al.\/}, {\it Nature\/} {\bf 467}, 68 (2010).

\bibitem{Gross:2021qg}
C.~Gross, W.~S. Bakr, {\it Nature Physics\/} {\bf 17}, 1316 (2021).

\bibitem{Vaucher:2008co}
B.~Vaucher, A.~Nunnenkamp, D.~Jaksch, {\it New Journal of Physics\/} {\bf 10},
  023005 (2008).

\bibitem{Yang:2017sd}
B.~Yang, {\it et~al.\/}, {\it Physical Review A\/} {\bf 96}, 011602 (2017).

\bibitem{Zupancic:2016up}
P.~Zupancic, {\it et~al.\/}, {\it Optics Express\/} {\bf 24}, 13881 (2016).

\bibitem{Li:2021hp}
M.-D. Li, {\it et~al.\/}, {\it Optics Express\/} {\bf 29}, 13876 (2021).

\bibitem{Boll:2016sa}
M.~Boll, {\it et~al.\/}, {\it Science\/} {\bf 353}, 1257 (2016).

\bibitem{Trotzky:2010ca}
S.~Trotzky, Y.-A. Chen, U.~Schnorrberger, P.~Cheinet, I.~Bloch, {\it Physical
  Review Letters\/} {\bf 105}, 265303 (2010).

\bibitem{Yang:2019ul}
Y.-M. Yang, {\it et~al.\/}, {\it Review of Scientific Instruments\/} {\bf 90},
  014701 (2019).

\bibitem{Xiao:2020gt}
B.~Xiao, {\it et~al.\/}, {\it Chinese Physics B\/} {\bf 29}, 076701 (2020).

\bibitem{Li:2021ah}
M.-D. Li, {\it et~al.\/}, {\it Review of Scientific Instruments\/} {\bf 92},
  083202 (2021).

\bibitem{Barenco:1995eg}
A.~Barenco, {\it et~al.\/}, {\it Physical review A\/} {\bf 52}, 3457 (1995).

\bibitem{Loss:1998qc}
D.~Loss, D.~P. DiVincenzo, {\it Physical Review A\/} {\bf 57}, 120 (1998).

\bibitem{Zhou:2022qg}
Y.~Zhou, {\it et~al.\/}, {\it npj Quantum Information\/} {\bf 8}, 99 (2022).

\bibitem{Briegel:2009mb}
H.~J. Briegel, D.~E. Browne, W.~D{\"u}r, R.~Raussendorf, M.~Van~den Nest, {\it
  Nature Physics\/} {\bf 5}, 19 (2009).

\bibitem{Tanamoto:2009eq}
T.~Tanamoto, Y.-x. Liu, X.~Hu, F.~Nori, {\it Physical review letters\/} {\bf
  102}, 100501 (2009).

\bibitem{zhang:2006mo}
C.~Zhang, S.~Rolston, S.~D. Sarma, {\it Physical Review A\/} {\bf 74}, 042316
  (2006).

\bibitem{Weitenberg:2011qc}
C.~Weitenberg, S.~Kuhr, K.~M{\o}lmer, J.~F. Sherson, {\it Physical Review A\/}
  {\bf 84}, 032322 (2011).

\bibitem{Yang:2020oo}
B.~Yang, {\it et~al.\/}, {\it Nature\/} {\bf 587}, 392 (2020).

\bibitem{Zhou:2022td}
Z.-Y. Zhou, {\it et~al.\/}, {\it Science\/} {\bf 377}, 311 (2022).

\bibitem{Yao:2022qm}
Z.~Yao, L.~Pan, S.~Liu, H.~Zhai, {\it Physical Review B\/} {\bf 105}, 125123
  (2022).

\bibitem{Cheng:2022tc}
Y.~Cheng, S.~Liu, W.~Zheng, P.~Zhang, H.~Zhai, {\it arXiv:2204.06586\/}
  (2022).

\bibitem{Halimeh:2022tt}
J.~C. Halimeh, I.~P. McCulloch, B.~Yang, P.~Hauke, {\it arXiv:2204.06570\/}
  (2022).

\bibitem{Bohrdt:2021eo}
A.~Bohrdt, L.~Homeier, C.~Reinmoser, E.~Demler, F.~Grusdt, {\it Annals of
  Physics\/} {\bf 435}, 168651 (2021).

\bibitem{Daley:2022pq}
A.~J. Daley, {\it et~al.\/}, {\it Nature\/} {\bf 607}, 667 (2022).

\bibitem{Ma:2011pa}
R.~Ma, {\it et~al.\/}, {\it Physical Review Letters\/} {\bf 107}, 095301
  (2011).

\bibitem{Preiss:2015sc}
P.~M. Preiss, {\it et~al.\/}, {\it Science\/} {\bf 347}, 1229 (2015).

\bibitem{Toth:2005ed}
G.~T\'oth, O.~G\"uhne, {\it Physical Review A\/} {\bf 72}, 022340 (2005).

\bibitem{Asadian:2016hw}
A.~Asadian, P.~Erker, M.~Huber, C.~Kl\"ockl, {\it Physical Review A\/} {\bf
  94}, 010301 (2016).

\end{thebibliography}
\bibliographystyle{Science}   % for Science mag.

%%%%%% Supplementary materials
\onecolumngrid
\vspace*{0.5cm}
\newpage
\begin{center}
    \textbf{METHODS AND SUPPLEMENTARY MATERIALS}
\end{center}
\vspace*{0.5cm}

\twocolumngrid
\incltocpage
\tableofcontents
\appendix
\setcounter{secnumdepth}{2}

\twocolumngrid
\setcounter{equation}{0}
\setcounter{figure}{0}
\makeatletter
\makeatother
\renewcommand{\theequation}{S\arabic{equation}}
\renewcommand{\thefigure}{S\arabic{figure}}
\renewcommand{\thetable}{S\arabic{table}}

% paragraph S01.
\section{Experimental system}

\begin{figure*}[!htb]
    \centering     %
    \includegraphics[width=140mm]{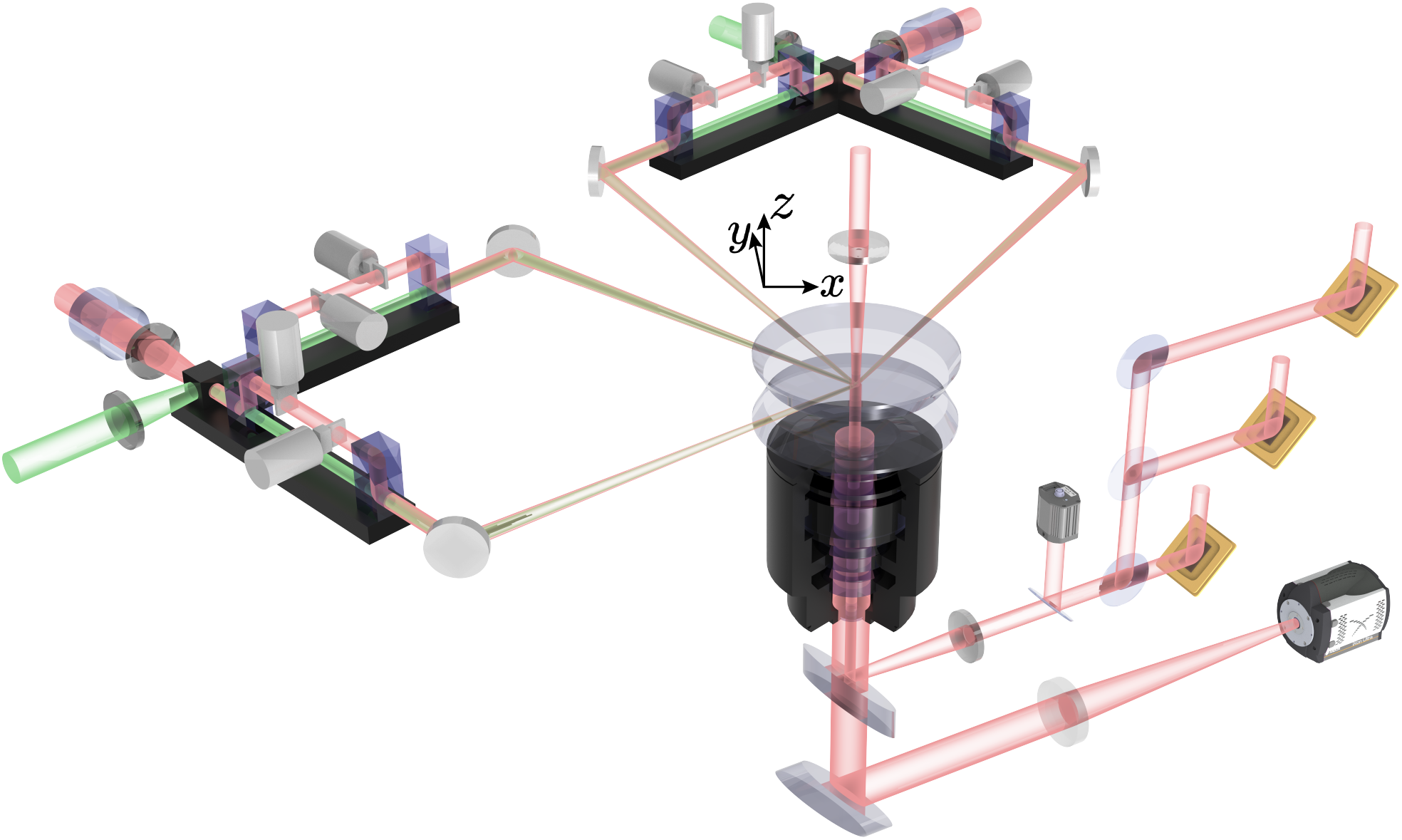}
    \caption{\textbf{Experimental setup}. 
    The 2D $^{87}\mathrm{Rb}$ atom gas is prepared in a single plane of vertical lattice along the $z$-direction, around $3~\mathrm{mm}$ above the upper surface of the super-polished vacuum window \cite{Xiao:2020gt}. And the vertical lattice along the $z$-direction is created by retroreflecting a laser beam at a wavelength of $1064~\mathrm{nm}$ on the same coated vacuum window. Both $x$- and $y$-directions contain a pair of short and long lattices formed by a blue-detuned laser and a red-detuned laser, respectively, with a lattice spacing of $a_{\mathrm{s}} = 630~\mathrm{nm}$ and $2a_{\mathrm{s}}$. Their intensity superimposes to form superlattice heterostructures whose relative phases can be flexibly tuned \cite{Li:2021hp}. Furthermore, electro-optical modulators (EOMs) in both the $x$- and $y$-directions of the red-detuned laser beam path can control the laser polarizations, thus forming the spin-dependent lattice potential mentioned later. An objective lens with a numerical aperture (NA) of $0.8$ is placed approximately $1~\mathrm{mm}$ below the viewport, which collects the scattered fluorescence photons from the atoms to record the corresponding position information. Light patterns for arbitrary potential tailoring and atom spin states manipulation generated by three independent digital micromirror devices (DMD1, DMD2, and DMD3) are projected to the atom cloud through the same objective lens.
    }
    \label{figureS1:setup}
\end{figure*}

\subsection{Experimental setup}
Our experimental setup is shown in Fig.~\ref{figureS1:setup}.
Along both $x$- and $y$-directions, we employ a blue-detuned optical ``short-lattice'' with a lattice spacing of $a_\mathrm{s} = 630~\mathrm{nm}$, created by crossing two laser beams of wavelength $\lambda_{\mathrm{s}} = 532~\mathrm{nm}$ at an angle $\alpha = 50^\circ$. In addition, ``long-lattice'' optical potentials along both $x$- and $y$-directions with twice the lattice constant of $a_\mathrm{l} = 2a_\mathrm{s}$ are created each by two laser beams of wavelength $\lambda_{\mathrm{l}} = 1064~\mathrm{nm}$ with the same optical paths as the corresponding ``short-lattice''. The intensity superposition of the ``short-lattice'' and the ``long-lattice'' generates the superlattice heterostructure \cite{Li:2021hp}. The resulting superlattice potential along the $x$- and $y$-directions reads
\begin{align}
    \label{equation:S1}
    V(x) & = V_{\mathrm{s},x} \cos^2(k_{\mathrm{s}} x) - V_{\mathrm{l},x} \cos^2(k_{\mathrm{l}} x + \theta_{x}) \, , \\
    V(y) & = V_{\mathrm{s},y} \cos^2(k_{\mathrm{s}} y) - V_{\mathrm{l},y} \cos^2(k_{\mathrm{l}} y + \theta_{y}) \, .
\end{align}
where $V_{\mathrm{s},x}$ ($V_{\mathrm{s},y}$) and $V_{\mathrm{l},x}$ ($V_{\mathrm{l},y}$) are the trap depths of the ``short lattice'' and the ``long lattice'' along the $x-$ ($y-$) direction, respectively. Here $k_{\mathrm{s}} = \pi / a_{\mathrm{s}}$ ($k_{\mathrm{l}} = \pi / a_{\mathrm{l}}$) is the wavenumber of the ``short lattice'' (``long lattice''), and $\theta_{x}$($\theta_{y}$) is the relative phase between the ``short lattice'' and the ``long lattice'' along the $x-$ ($y-$) direction. Two electro-optical modulators (EOMs) are applied in the $x$- and $y$-directions to control the laser polarization of the ``long-lattice'', respectively. Therefore, we can further use the combination of the ``long-lattice'' and the corresponding ``short-lattice'' to generate a state-dependent superlattice potential \cite{Yang:2017sd}, which offers the opportunity to manipulate the atomic spin states in parallel. Our platform equips with a quantum gas microscope. The atomic distribution after a parity projection is recorded by the site-resolved fluorescence imaging \cite{Bakr:2009aq,Sherson:2010sa}. Our quantum gas microscope can not only provide the technique of the site-resolved optical detection but also the technique of the site-resolved manipulation. The precise manipulation of atoms is implemented by equipping three digital micromirror devices (DMD1, DMD2, and DMD3), as shown in Fig.~\ref{figureS1:setup}.

\subsection{2D cloud preparation}
Our experiment starts with a 2D Bose-Einstein condensate of $^{87}$Rb atoms in the $5\mathrm{S}_{1/2}$ $\ket{F=1,m_\mathrm{F}=-1}$ state prepared in a single plane of a vertical lattice along the $z$-direction, which has been described in  \cite{Xiao:2020gt}. The laser for the vertical lattice has a wavelength of $1064~\mathrm{nm}$ and a beam waist $\sim 103~\mu \mathrm{m}$. To further cool the atoms, we apply an extra dimple trap to atoms at the center of the cloud. The wavelength of the dimple beam is $850~\mathrm{nm}$, and the beam waist is $\sim 10~\mu \mathrm{m}$. We then apply a magnetic field gradient of $15~\mathrm{G/cm}$ along the horizontal direction to remove the atoms outside the dimple trap. After that, we perform the evaporation cooling by lowering the power of the dimple beam to a suitable final value. Adjusting the final depth of the dimple trap enables us to finely tune the total number of atoms after cooling. Finally, we turn off the magnetic field and lower the dimple trap gradually to release the atom cloud into a single antinode of the vertical lattice along the $z$-direction.

\begin{figure*}[htb]
    \centering     %
    \includegraphics[width=150mm]{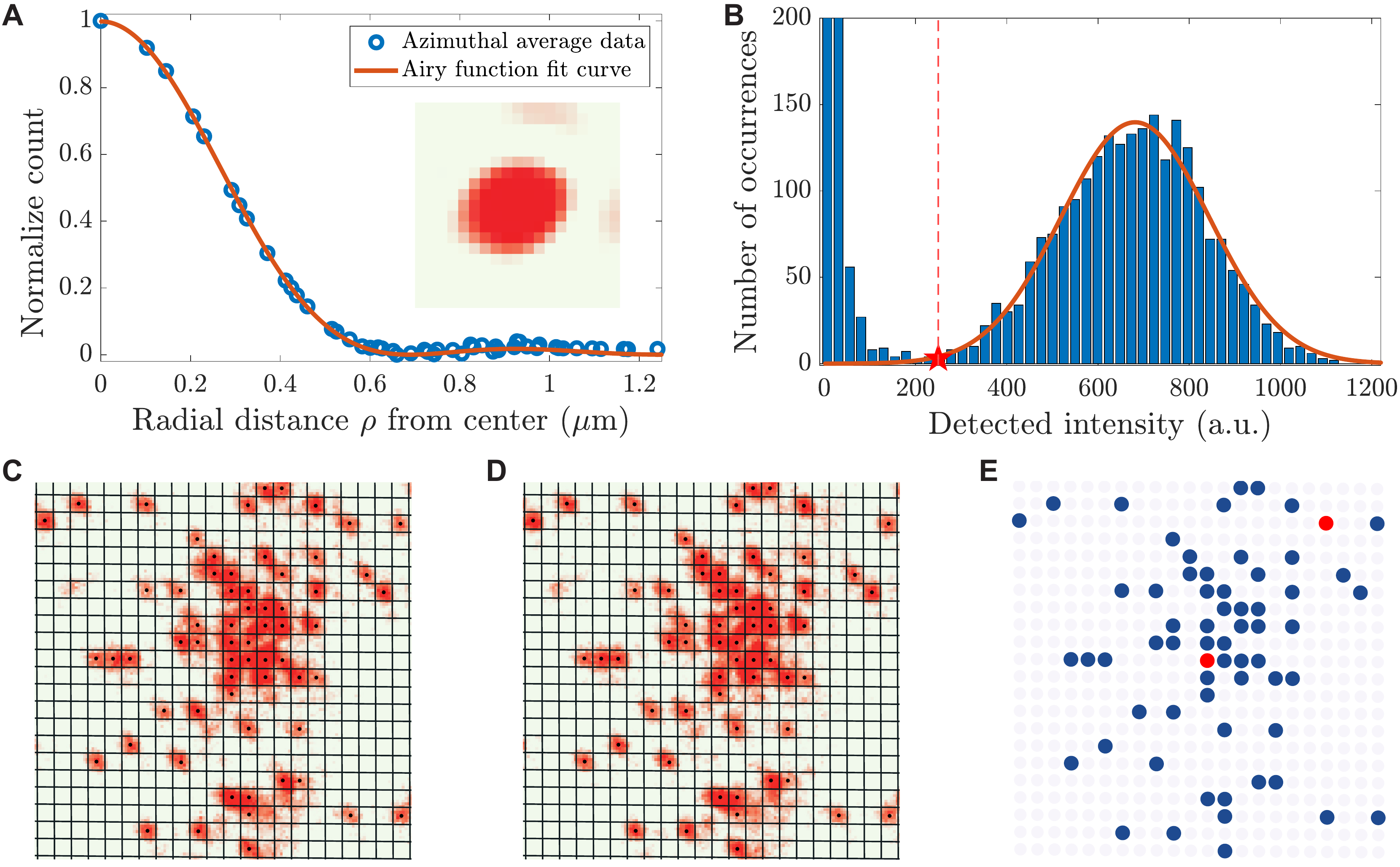}
    \caption{\textbf{Site-resolved Imaging}.
    (\textbf{A}) Measured point-spread function (PSF) of our imaging system. The inset image is obtained by averaging the signals of 103 isolated single atoms. The azimuthal average of the data (blue circles) and the corresponding curve fitted by the Airy function (red line) give a resolution of $577(5)~\mathrm{nm}$, which is expected for our imaging system with a numerical aperture NA = 0.8. (\textbf{B}) Histogram of the brightness distribution of lattice sites for an exposure time of $0.5~\mathrm{s}$. The red star and the red dashed line mark the threshold of distinguishing an empty site from an occupied one. The solid curve is a Gaussian fit of the right peak, which determines the threshold. The left part corresponds to empty sites (background subtracted), and the right part comes from occupied sites. (\textbf{C} and \textbf{D}) are two consecutive images of the same dilute atom cloud. (\textbf{E}) The reconstructed atom distribution of C and D. The blue spots represent atoms presented in both images and the red spots mark the loss events (no hopping events have been recorded). As shown in E, two atoms are lost in the second image, which gives a loss rate of $3.13\%$.}
    \label{figureS2:siteResImg}
\end{figure*}

\subsection{Single-site-resolved imaging}
To perform the single-site-resolved fluorescence imaging, we first increase the lattice depth to $\sim 300~\mu\mathrm{K}$ along all three dimensions immediately after each experimental sequence, and then illuminate the atom cloud with an optical molasses for $0.5~\mathrm{s}$. The scattered fluorescence photons are collected by the objective lens and recorded by an electron-multiplying charge-coupled device (EMCCD) camera. To characterize the capability of our imaging system, we measure the point spread function (PSF) using hundreds of isolated atoms. Using the Airy function to fit the azimuthal average of the PSF from the averaged single atom signal, we obtain an optical resolution of $577(5)$ nm (full-width at half-maximum, FWHM) at the wavelength of $780$ nm, as shown in Fig.~\ref{figureS2:siteResImg}A. We develop a deconvolution algorithm to reconstruct the original site-occupation with a fidelity of 99.6(3)\%. We determine the presence of an atom when the number of photons collected in a predefined sensor region on the camera exceeds a threshold value, as shown in Fig.~\ref{figureS2:siteResImg}B. However, background gas collisions or noise-induced heating during imaging may cause atoms to be lost from the trap or hop to other lattice sites. Therefore, we further evaluate the fidelity of our imaging system by measuring the hopping and loss rates during the fluorescence imaging process. We take two consecutive images of the atom cloud and compare their reconstructed site occupations. The exposure time of each image is $0.5$ s, and the interval between two images is also $0.5$ s. We count the sites that are empty in the first image and occupied in the second image as hopping events. And we consider sites occupied in the first image and empty in the second image as loss events. For the optimized parameters used in the experiment, we achieve a loss rates of $\sim 3.1\%$ and a hopping rates of $\sim 0.1\%$ during the $0.5$ s exposure time of the imaging process, as shown in Fig.~\ref{figureS2:siteResImg}, C to E. Taking into account the loss rate and hopping rate, we achieve a total efficiency of $96.8(4)\%$ for identifying the atoms at a given lattice site. The loss rate will be included in the correction of the occupation number.

\section{Calibrations}

\subsection{Lattice depths and superlattice phases}

\begin{figure*}[htb]
    \centering     %
    \includegraphics[width=175mm]{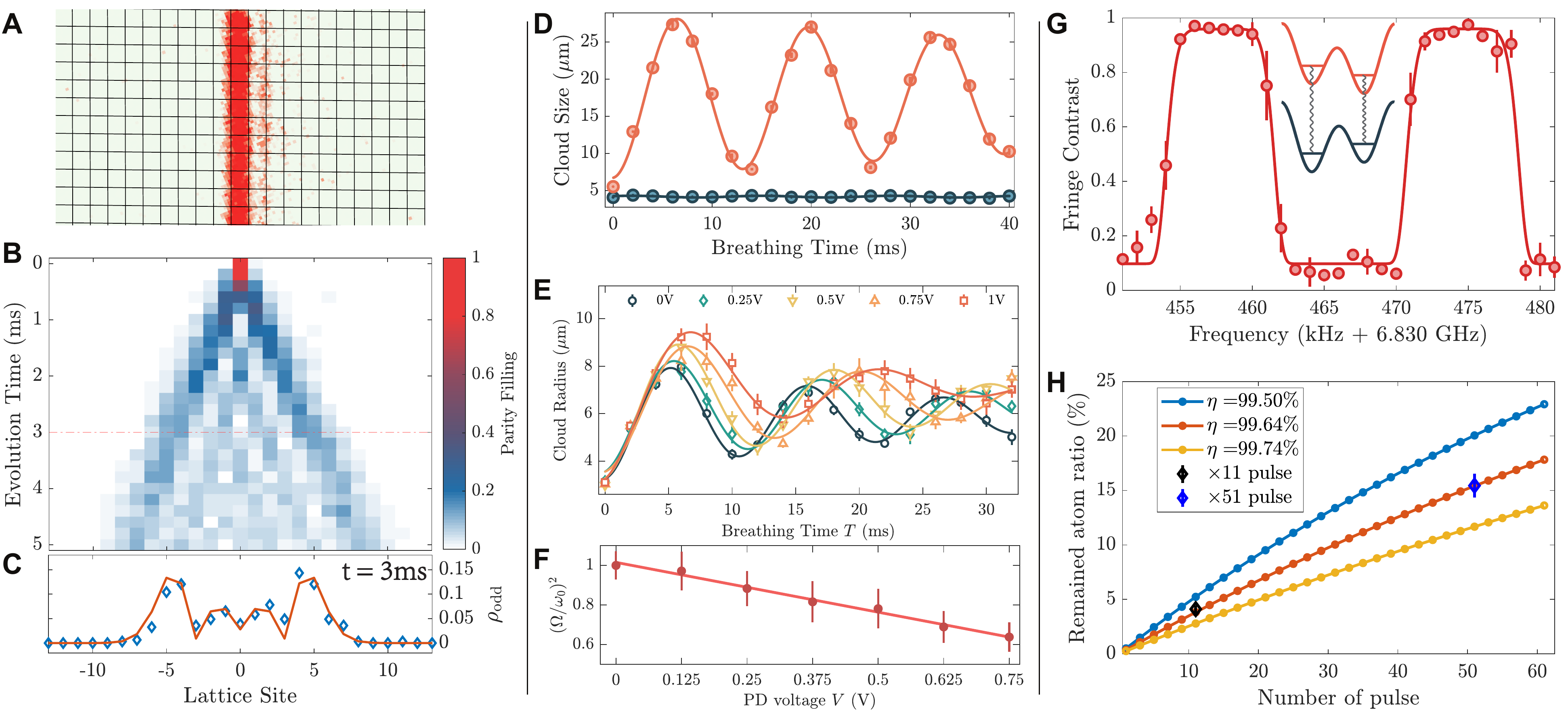}
    \caption{\textbf{Calibrations}. 
    (\textbf{A}) The exemplary images of one single-chain after the single-site-resolved addressing operation. (\textbf{B}) Time-resolved tracing of the averaged density distribution of a single-atom coherent quantum walk shows an apparent linear expansion in time. (\textbf{C}) The averaged density distribution at an evolving time $t=3~\mathrm{ms}$ with the corresponding fitted curves based on Eq.~\ref{equation:QWs}. (\textbf{D}) Overall harmonic trapping frequency calibration in the presence of the ``short lattice'' along the $x$-direction and the vertical lattice along the $z$-direction. In this case, the atomic cloud can only exhibit a significant breathing phenomenon in the direction of the 1D tubes, i.e. the $y$-direction. The dots represent the observed atomic cloud size with different breathing oscillation times, and the solid curves indicate the sinusoidal fit to the experimental data. The overall harmonic trapping frequency can be extracted from the sinusoidal fitted result. (\textbf{E}) Observed atom cloud size during breathing-mode dynamics with different DMD power. The solid curves are the fitting results from the damped sinusoidal function, which can give us the corresponding oscillation frequencies. (\textbf{F}) The fitted breathing mode frequency ratios$(\Omega/\omega_{0})^{2}$ and the DMD power have a linear relation. (\textbf{G}) Two distinct plateaus with a 16.8(2)$~\mathrm{kHz}$ separation are observed, corresponding to the frequency difference of microwave pulses used to individually address the left and the right lattice sites in the site-dependent superlattices. (\textbf{H}) The black and blue diamond dots represent the residual atom ratio after 11 and 51 microwave pulses, respectively. The solid curves with circles are the theoretical expected residual ratio with corresponding efficiencies. Error bars denote the s.e.m in (C to E, G and H), and are smaller than the markers if hidden. Error bars represent the fitting error in F.
    }
    \label{figureS3:calibrationS}
\end{figure*}

\noindent \textbf{Lattice depth}.
We use lattice modulation spectroscopy to calibrate the lattice depths of the ``short lattice'' and the ``long lattice'' along the $x$- and $y$-directions and the vertical lattice along the $z$-direction. After preparing the 2D atomic cloud, we ramp up the lattice to the desired depth in $60~\mathrm{ms}$ and then add a sinusoidal modulation to the lattice depth at a given frequency with an amplitude of $6\%$ for $2$ ms. We then employ the band mapping technique by turning off the lattice potential at 300 $\mu$s, letting atoms freely expand for 10 ms, and performing fluorescence imaging after suddenly freezing the atoms. We repeat the above procedure with varied modulation frequencies. Based on the measured width of the atom cloud after modulation, we use a Lorentzian fit to extract the resonant frequency, which corresponds to the energy difference between the $s$ and the $d$ bands. The lattice depth is then determined by comparing the extracted resonant frequency with theoretical band structure calculations.

\vspace{5pt}
\noindent\textbf{Superlattice phase}.
We calibrate the superlattice phase $\theta_{x}$ (or $\theta_{y}$), which is controlled by the galvanometers, with the same method described in our previous works \cite{Li:2021hp}. After preparing the 2D atomic cloud, we adiabatically ramp up the ``long lattice" along the $x-$ (or $y-$) direction to a depth of $20E_\mathrm{r}$, where $E_{\mathrm{r}} = h^2 / (8 m_{\mathrm{Rb}} a_\mathrm{s}^2)$ is the recoil energy and $m_{\mathrm{Rb}}$ is the mass of the $^{87}\mathrm{Rb}$ atom. We then gradually increase the depths of the ``short lattices" along both $x$ and $y$-directions to $60E_\mathrm{r} $ to separate the atoms into series of double wells along the $x-$ (or $y-$) direction. After that, we immediately freeze the atoms and record the distribution using fluorescence imaging. We use a Lorentzian fit to extract the contrast $I = \left|(n_\text{L}-n_\text{R})/(n_\text{L}+n_\text{R})\right|$, where $n_\text{L}$ and $n_\text{R}$ are the averaged parity of the occupation number on the left and right sides of double wells, respectively. The dips (or the peaks) positions correspond to balanced double-well structures, which determine the relative phase between the ``short lattice" and the ``long lattice" as $\theta_{x} = 0$ (or $\theta_{y} = 0$).

\subsection{Hubbard parameters}
\vspace{5pt}
\noindent \textbf{Hubbard parameters $U$}.
We calibrate the on-site interaction strength $U$ using the method in Ref.~\cite{Ma:2011pa}. After preparing copies of near defect-free 1D Mott insulator chains along the $x$- (or $y$-) direction, we employ a linear potential along the same direction using a magnetic field gradient. Then we decrease the lattice depth to the value we use in the experiments and modulate the depth of our ``short lattice" along the $x-$ (or $y-$) direction. We measure the parity occupations as a function of the modulation frequency. The resulting curve shows two minima at frequencies $E_{\mathrm{tilt}} \pm U$. The obtained $E_{\mathrm{tilt}}$ is the energy difference of neighboring sites due to the linear tilt.

\vspace{5pt}
\noindent \textbf{Hubbard parameters $J$}.
We calibrate the tunneling strength $J$ through the 1D quantum walk \cite{Preiss:2015sc}. This procedure starts with copies of near defect-free 1D Mott insulator chains. We projected a particular light pattern via DMD2 to select only one single-chain along the $y$-direction, as shown in Fig.~\ref{figureS3:calibrationS}A. We keep the $y$ ``short lattice'' at $60E_{\mathrm{r}}$ to freeze the tunneling along the $y$-direction. The coherent single-atom quantum walk is initiated by quickly ramping down the depth of the $x$ ``short lattice'' to a desired value in $300~\mu \mathrm{s}$. After letting the system free evolve for various times $t$, we detect the single atom position via the fluorescence imaging. Fig.~\ref{figureS3:calibrationS}B shows the time-resolved tracing of the averaged density distribution over eight adjacent rows along the $y$-axis in the central region. The measured density distribution $p(t)$ expands linearly versus time $t$, showing a ballistic transport originated from the coherent interference of all possible paths, agreeing excellently with the theoretical expectation, which can be expressed as
\begin{equation}
p_i(t) = \left|\mathcal{J}_{i}\left(\frac{2 J}{\pi E} \sin (\pi E t / h)\right)\right|^{2}
\label{equation:QWs}
\end{equation}
\noindent where, $h$ is the Planck’s constant, $J$ is tunneling strength along the $x$-axis, $i$ denotes the distance to the initial atom position in the unit of lattice sites, $\mathcal{J}_{i}$ is the $i^{\mathrm{th}}$-order Bessel function of the first kind, and $E$ represents the possible existing energy tilt between the adjacent lattice sites.

\subsection{Harmonic trapping frequency}
\vspace{5pt}
We calibrate the global harmonic confinement resulting from the Gaussian profile of lattice beams by observing breathing mode oscillations. After the preparation of a 2D atom cloud, we load the atoms into the trap formed by lasers of interest and the dimple trap \cite{Xiao:2020gt}, then turn off the dimple trap suddenly to drive the breathing mode along the direction normal to the lattice for a period time. Finally, we detect the atomic distribution with fluorescence imaging. To extract the oscillation frequency, the cloud sizes with different oscillation times are fitted using a sinusoidal function, with the fitted frequency corresponding to twice the harmonic trap frequency. Fig.~\ref{figureS3:calibrationS}D shows the experimentally observed breathing oscillation signal in the presence of ``short lattice'' along the $x$-direction and vertical lattice along the $z$-direction .

\subsection{DMD potential calibration}
As shown in Fig.~\ref{figureS1:setup}, our platform equips three digital micromirror devices (DMD1, DMD2, and DMD3, respectively), which provide the capability to precisely manipulate atoms. DMD1 and DMD3 are designed to generate repulsive potentials to compensate for spatial inhomogeneities or to create isolating walls with blue-tuned light at a wavelength of $750~\mathrm{nm}$. DMD2 is designed to realize local atomic spin addressing capability in an arbitrary configuration at the single-site-resolved level. We use a circularly polarized laser at a wavelength of 787.55 nm, a magical wavelength between the $D1$ and $D2$ transitions \cite{Weitenberg:2011ss}.

\vspace{5pt}
\noindent \textbf{DMD1 and DMD3.} 
To calibrate the potential projected by the DMD1 (or DMD3), we utilize an anti-Gauss pattern with a radius of 17 $\mu$m to compensate for the overall harmonic trapping potential originating from the lattice beams. The overall harmonic trap frequency, which can be tuned by varying laser power incidents on the DMD1 (or DMD3), is measured by the breathing mode oscillations method mentioned above. From the fitting results in Fig.~\ref{figureS3:calibrationS}E, the breathing mode frequency $\Omega$ can be extracted, while $\omega_{0}$ is the breathing mode frequency in the absence of DMD projection. We use a linear function to fit the power of the laser beam applied to the DMD which is characterized by the voltage $V$ read from the photodiode. The ratio $(\Omega/\omega_{0})^{2}$ is extracted from the fitting result above to obtain the depth of the potential projected through the DMD1 (or DMD3), as shown in Fig.~\ref{figureS3:calibrationS}F.

\noindent \textbf{DMD2.} 
To calibrate the potential projected by the DMD2, we directly observe the differential light shift between $\ket{\uparrow}$ and $\ket{\downarrow}$ states via microwave (MW) spectroscopy. We begin by preparing the atomic cloud in the $\ket{\uparrow}$ state, and at this moment, the resonance frequency between $\ket{\uparrow}$ and $\ket{\downarrow}$ states is $\Omega_0$, captured by the MW spectroscopy. Then we project a flattop potential onto the atoms with DMD2 and scan the MW frequency to flip the atoms from the $\ket{\uparrow}$ state to the $\ket{\downarrow}$ state, followed by a resonance laser beam to remove the atoms in the $\ket{\uparrow}$ state. We can obtain the new resonance frequency $\Omega$ once the remaining atom number reaches the maximum. We can extract the projected potential from the frequency difference $\delta = \Omega_0 - \Omega$ according to the above procedures.

\subsection{Spin-dependent effect and microwave efficiency}
\vspace{5pt}
Parallel atomic spin state addressing is achieved by introducing spin-dependent effects on the superlattice \cite{Yang:2017sd}. We perform the parallel spin state addressing by setting the depths of the ``short lattice'' and ``long lattice'' to $\sim 37.5~E_{\mathrm{r}}$ and $\sim 96~E_{\mathrm{r}}$, respectively. Meanwhile, we fix the bias field along the $z$-direction at approximately $2~\mathrm{Gauss}$ and tune the phase of the electro-optical modulator (EOM) to $\sim 50$ degrees. Therefore, we create an $16.8~\mathrm{kHz}$ energy splitting between the two hyperfine states (denoted by $\ket{\downarrow} = \ket{F=1,m_{\mathrm{F}}=-1}$ and $\ket{\uparrow} = \ket{F=2,m_{\mathrm{F}}=-2}$) of the odd and even sites in each double well, as shown in Fig.~\ref{figureS3:calibrationS}G. We then employ a Landau–Zener (LZ) crossing process to transfer the atoms on odd (or even) sites from $\ket{\downarrow}$ state into $\ket{\uparrow}$ state \cite{Xiao:2020gt}. To calibrate the fidelity of the spin-flip operation, we apply a multi-pulse sequence to the atomic cloud. We deduce from the residual atom ratio after 11 and 51 pulses that the fidelity of flipping odd sites without affecting even sites is 99.64(4)\%, as shown in Fig.~\ref{figureS3:calibrationS}H.

\subsection{Calibration of the $\sigma_x^{\otimes N} \sigma_z^{\otimes N}$ measurement}
To characterize the property of the generated multipartite entangled state specified in the main text, we need to obtain the spin correlation under the $\sigma_x \sigma_z$ basis, as illustrated in Fig.~\ref{figure4:2DEntangledS}, B and C. The basic idea of such an implementation combines the spin-dependent superlattice, which introduces an energy bias between the adjacent lattice sites in each double well, and a Rabi flopping microwave (MW) $\pi/2$ pulse on the atom in either the left or right site.

\begin{figure}[htb]
    \centering     %
    \includegraphics[width=85mm]{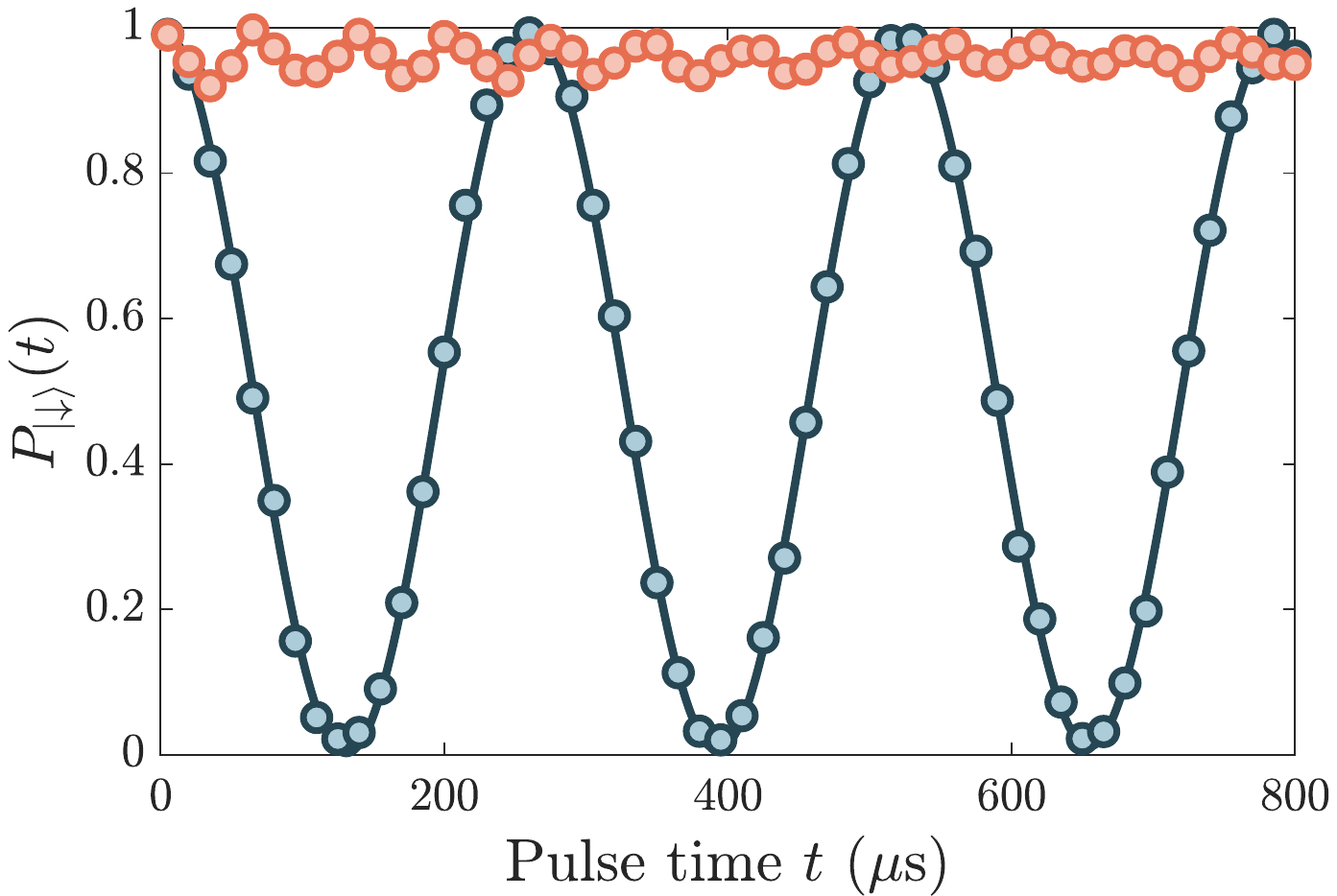}
    \caption{\textbf{Driven Rabi oscillation in  spin-dependent superlattices}. 
    The circles represent the measured time-resolved tracing of the probabilities of the $\ket{\downarrow}$ spin state in the left (blue) and right (orange) sites of each double well, respectively. The solid lines represent the fitting curves with the sine function.
    }
    \label{figureS4:XZM}
\end{figure}

We calibrate the Rabi frequency of the MW pulse and the energy difference induced by the spin-dependent effect as follows. After preparing the alternately arranged near defect-free Mott insulator chains along the $x$-direction, all atoms are in the $\ket{\downarrow}$ spin state. We first turn on the spin-dependent superlattice with given lattice depths along the $x$-direction and then employ a driven Rabi pulse  in resonance with the atoms on the left side of each double well. With varying driving times, we conduct the Stern-Gerlach-type measurement (see the following part) to record the spin state at each lattice site. The measured time-resolved tracing of the probabilities of the $\ket{\downarrow}$ spin state in the left and right sites of each double well, respectively, is shown in Fig.~\ref{figureS4:XZM}. The fitted results show that the $\pi/2$ rotation pulse acting on the left (or right) site lasts for $t_{\pi/2} \approx 65 ~\mu$s. Furthermore, the spin state on the opposite side of the identical double well remains in the $\ket{\downarrow}$ spin state with a probability of 99.71\% after the $\pi/2$ rotation pulse.

The Rabi oscillation curves, as illustrated in Fig.~\ref{figureS4:XZM} above and Fig.~\ref{figure1:stagC}E in the main text, exhibit the excellent coherence property of our qubit arrays. This remarkable coherence is thanks to the advancement of low-noise homemade current sources accompanied by active magnetic field compensation \cite{Yang:2019ul,Xiao:2020gt}, as well as the adoption of extremely low-noise lattice lasers that also actively suppress intensity noise \cite{Li:2021ah}.

\section{Staggered-immersion cooling}

We employ the recently demonstrated staggered-immersion cooling method to prepare a nearly unity-filling Mott insulator as the basis for performing subsequent studies on the generation and determination of quantum multiparticle entangled states \cite{Yang:2020ca}. The experiment begins with a 2D quantum gas in a single layer of the vertical lattice along the $z$-direction \cite{Xiao:2020gt}. We drive the phase transition by ramping up the depth of the ``short-lattice'' along both $x$- and $y$-directions linearly from $2.5E_{\mathrm{r}}$ to $~20E_{\mathrm{r}}$ across the critical point in $80~\mathrm{ms}$. Where $E_{\mathrm{r}} = h^2 / 8 m a_\mathrm{s}$ is the recoil energy, $h$ is the Planck's constant, $a_\mathrm{s}$ represents the lattice spacing of the ``short lattice'' and $m$ denotes the atomic mass. Simultaneously with the lattice loading, a repulsive potential is ramped up, compensating for the harmonic confinement originated by the Gaussian envelope of lattice beams. We also superimpose a ``long-lattice'' along the $x$- (or $y$-) direction before the phase transition to construct the staggered system, as shown in the inset of Fig.~\ref{figure1:stagC}A. The staggered system has an average site occupation of $\sim 0.75$. As we increase the $U/J$, where $U$ is the on-site interaction and $J$ is tunneling strentgh, the atoms undergo a phase transition to the Mott insulator phase in the deep lattice sites ($\epsilon_i = 0$), while in the shallow lattice sites ($\epsilon_i = U / 2$), the atoms stay in the superfluid phase. Since the gapped Mott insulators are submerged into the gapless superfluid reservoirs, the induced excitations due to nonadiabatic drive in the phase transition will mainly occur in the reservoirs. After the phase transition process, we remove all atoms in the superfluid reservoirs via site-dependent addressing \cite{Yang:2017sd}. The in-situ atomic parity distribution of the retained sample systems is then recorded by the single-site resolved fluorescence imaging.

\begin{figure}[htb]
    \centering     %
    \includegraphics[width=85mm]{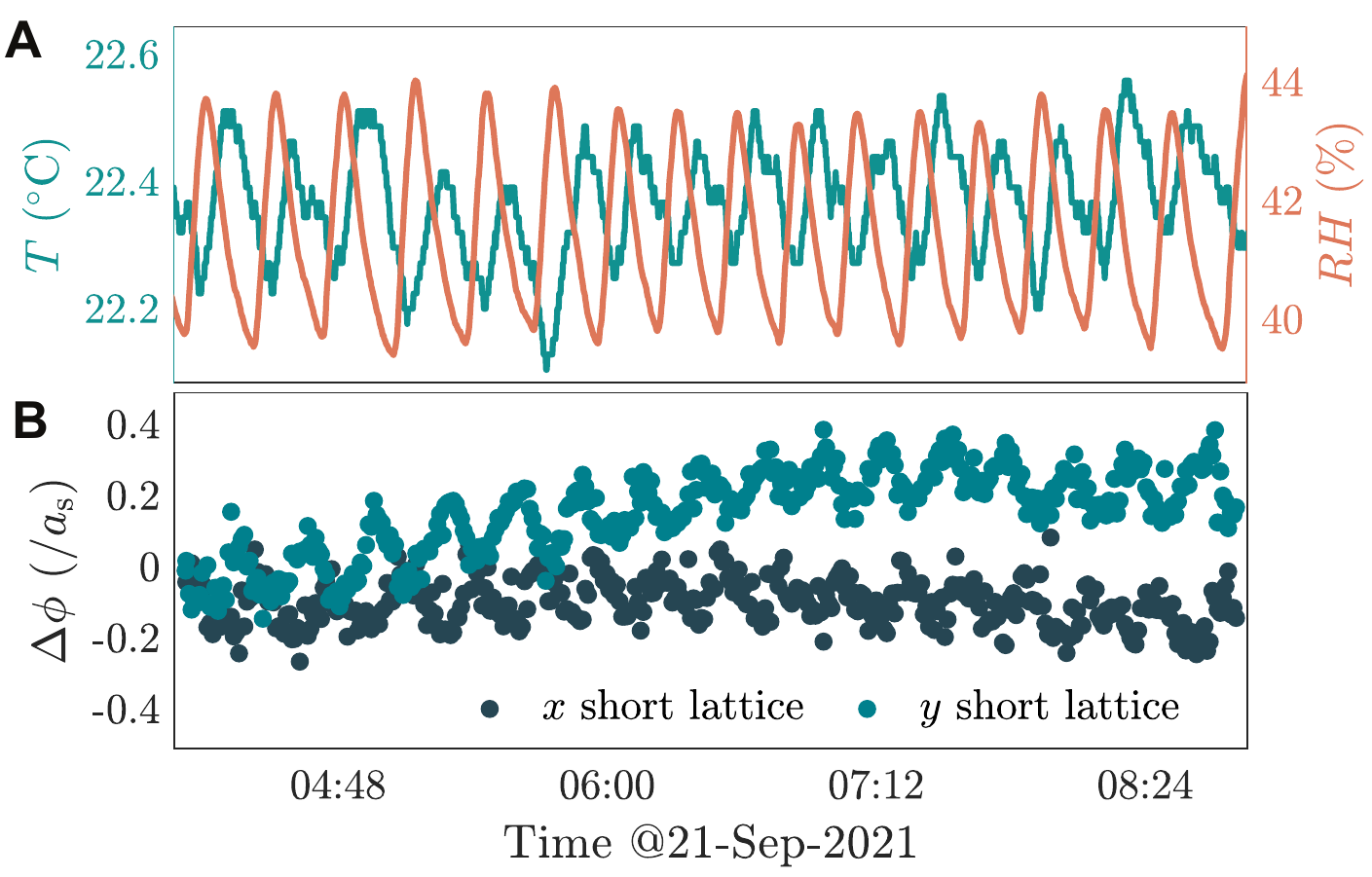}
    \caption{\textbf{Stability of the environment and the lattice phases}. 
    (\textbf{A}) Monitoring of the environment temperature ($T$) and humidity ($RH$) changes inside the laboratory. (\textbf{B}) Tracing of the phase of short lattices along both $x$- and $y$-directions, in the unit of lattice site $a_{\mathrm{s}}$. Error bars represent the s.e.m.
    }
    \label{figureS5:stagC}
\end{figure}

Variations in the initial state and changes in environmental conditions can cause a discrepancy in the cooling effect and deviation from theoretical expectations during the experimental repetition, which is manifested by fluctuations in the filling of the atomic number in the ROI region in practical experiments as shown in Fig.~\ref{figure1:stagC}A. First, as shown in Ref.~\cite{Yang:2020ca}, the initial atom filling rate of the whole system, as well as the corresponding entropy before the staggered-immersion cooling process, are critical to the final cooling effectiveness. Second, fluctuations in the environmental temperature and humidity within the experimental platform cause a shift in the pointing of the laser beam (see Fig.~\ref{figureS5:stagC}), which affects the final atom occupancy during the different experimental repetitions. Furthermore, due to the finite vacuum-limited lifetime, the background gas may collide with the atoms during the experiment, causing atoms to be lost from the trap. As a result, when estimating the average occupancy of the sample systems, we excluded those repetitions where the center of mass of the atomic cloud shifted by more than four lattice constants and where the number of atoms was less than half of the total number of lattice sites in the ROI. Of course, the effect of atomic loss during fluorescence imaging also contributes to the filling rate fluctuations. The loss rate will be included in the correction of the occupation number. Therefore, we deduce a filling factor of 99.2(2)\% for the prepared Mott insulator state in the ROI after the staggered-immersion cooling process (the value in parentheses represents s.e.m).

\section{1D and 2D spin-resolved detection}

We specifically developed fully spin-resolved detection techniques for both 1D and 2D systems in our experiments.

\subsection{1D spin-resolved detection}
We utilize alternately arranged unoccupied auxiliary lattice sites combined with a magnetic field gradient to conduct a Stern-Gerlach-type detection to extract the spin configuration of a 1D spin chain (or an isolated double-well). As illustrated in Fig.~\ref{figureS7:superExch}B, we perform the following operation to read out the final spin configuration after the atomic superexchange dynamics in the isolated double-well along the $y$-direction. We first transfer them into the ``long lattice'' along the $x$-direction after modifying the relative phases between the long and short lattices to overlap the intensity minima of the lattices. Following the turning off of the ``short lattice'' along the $x$-direction, we re-adjust the relative phase between the long and short lattices to match the position of the intensity maxima of the lattices. Then we turn on the magnetic field gradient so that the potential minima experienced by the two spin states separate in opposite directions along the $x$-direction. After that, we adiabatically ramp up the ``short lattice'' to complete the separation of the two atoms with distinct spin states into the two different sites of the local double well. Subsequently, we turn off the magnetic field gradient and perform fluorescence imaging to record the occupancy of the atoms. The spin states of the atoms can be distinguished simultaneously by their final locations.

\subsection{2D spin-resolved detection}
The above-mentioned Stern-Gerlach-type technique no longer directly applies to 2D qubit arrays, as there are no longer alternatively arranged unoccupied auxiliary lattice sites. Therefore, we develop a new approach to achieve the 2D spin-resolved detection, which employs the state-dependent atom transport technique. As illustrated in Fig.~\ref{figure4:2DEntangledS}A in the main text, we conduct the following procedures to read out the spin configuration in a 2D plaquette system along the $y$-direction. We first transfer the two atomic chains into two adjacent double wells along the $x$-direction. Then we shine a state-dependent addressing beam to pin those atoms in the $\ket{\uparrow}$ state with the DMD2. After that, we change the phase of the long-lattice potential to transport those atoms in the $\ket{\downarrow}$ state to the originally unoccupied auxiliary sites before performing the fluorescence imaging. The original spin state of the atoms can also be determined simultaneously with their final spatial locations.

\section{State preparation and measurement (SPAM) correction}

\begin{figure}[htb]
    \centering     %
    \includegraphics[width=85mm]{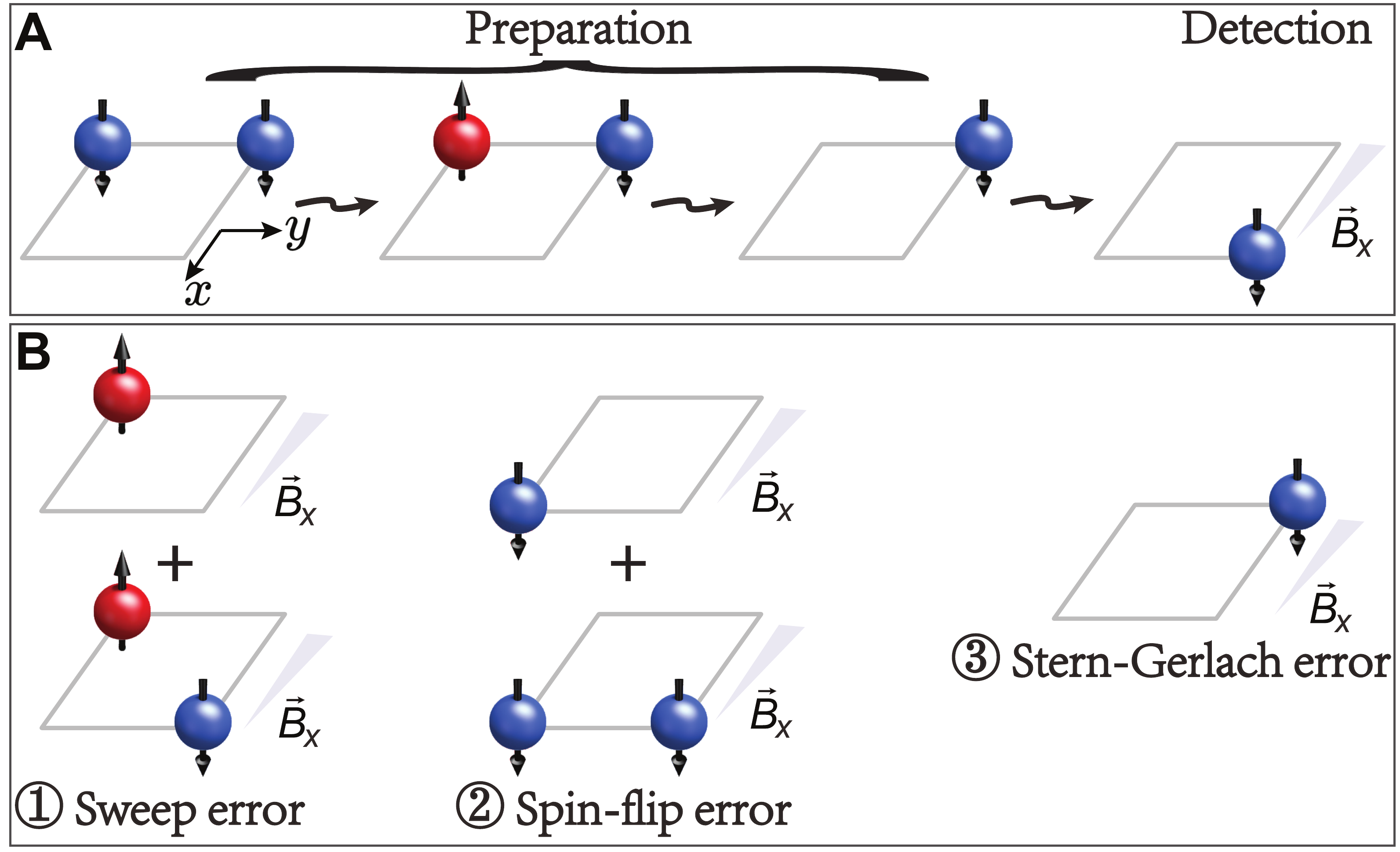}
    \caption{\textbf{SPAM evaluation and error syndromes in 1D spin-resolved detection}. (\textbf{A}) Cartoon sketch of the SPAM evaluation. (\textbf{B}) Cartoon sketch of the error syndromes in the SPAM described in the text.
    }
    \label{figureS6:spam}
\end{figure}

\subsection{SPAM in 1D spin-resolved detection}
Each experimental repetition involves the preparation of the initial state, the operation sequence, and the final state detection, whether in the study of atomic spin superexchange or the generation of entangled atom pairs and chains. Although post-selection rules can help us drastically mitigate the impact of atom loss during fluorescence imaging, there are three major error syndromes in state preparation and measurement (SPAM), which cause the misidentification of the final spin configuration, summarized as follows.
\begin{itemize}
  \item \textbf{Sweep error}. After the staggered-immersion cooling process, we employ an LZ-type microwave pulse followed by a resonant laser pulse to remove the atoms from the reservoir (shallow lattice sites). This procedure yields certain residual atoms due to the laser pulse efficiency, called the  \textbf{Sweep error}.
  \item \textbf{Spin-flip error}. We used an LZ-type microwave pulse to transfer the atomic spin state on the odd (left) side of the double wells from the $\ket{\downarrow}$ spin state to the $\ket{\uparrow}$ spin state. The fluctuation of the external magnetic field and improper selection of the pulse duration time restrict the pulse efficiency, causing the so-called \textbf{Spin-flip error}.
  \item \textbf{Stern-Gerlach error}. We use a Stern-Gerlach-type technique to separate the atoms with different spin states into opposite lattice sites in each plaquette to detect the final spin configurations of the atoms after spin operations. However, the procedures above may lead to some mistakes due to superlattice phase fluctuations and other experiment imperfections. We then attributed it to the \textbf{Stern-Gerlach error}.
\end{itemize}

% error budget table
\begin{table}[!htp]
\centering
\caption{Error budget}
\label{table:errorbudget}
\begin{tabular}{ p{1.5cm}|p{3.0cm}|p{2.5cm} }
\hline
 & Error syndromes & Error rate \\ 
\hline
\multirow{3}{2em}{1D detection} & Sweep error & 0.07\% $\pm$ 0.02\% \\ 
 & Spin-flip error & 0.23\% $\pm$ 0.04\% \\ 
 & Stern-Gerlach error & 0.50\% $\pm$ 0.02\% \\
\hline
\multirow{4}{2em}{2D detection} & $\ket{\uparrow}$ in $1^{\mathrm{st}}$ row & 0.87\% $\pm$ 0.35\% \\
 & $\ket{\downarrow}$ in $1^{\mathrm{st}}$ row & 2.52\% $\pm$ 0.46\% \\
 & $\ket{\uparrow}$ in $2^{\mathrm{nd}}$ row & 1.25\% $\pm$ 0.28\% \\
 & $\ket{\downarrow}$ in $2^{\mathrm{nd}}$ row & 1.27\% $\pm$ 0.28\% \\
\hline
\end{tabular}
\end{table}

We conducted the following experimental test to calibrate the corresponding error rates for the above error syndromes, as shown in Fig.~\ref{figureS6:spam}A. This procedure starts from the preparation of a series copies of near unit-filled Mott insulators arranged alternately along the $y$-direction. Then, we turn on the spin-dependent superlattice along the $y$-direction. After that, we apply an LZ-type microwave pulse to flip the spin state at the odd (left) sites of each double-well along the $y$-direction in the presence of a spin-dependent superlattice. Next, we remove those spin-flipped atoms with a resonant laser pulse. At last, we carry out the aforementioned Stern-Gerlach-type method to detect the final spin states in each plaquette. These operations mentioned above contain the three primary error syndromes, as illustrated in Fig.~\ref{figureS6:spam}B. By analyzing the results of the above experiments, we can extract the probability of error for the corresponding error symptoms. The corresponding error budget is summarized in TAB.~\ref{table:errorbudget}.

\subsection{Detection efficiency in 2D spin-resolved detection}
Cross-talking between different spin states, as well as accidental atom-hopping events during the detection procedure, decreases the efficiency of the aforementioned 2D spin-resolved detection scheme. To calibrate the detection efficiency, we repeat the spin-resolved detection procedures to a given initial state $\ket{\uparrow,\downarrow;\downarrow,\uparrow}$, and then take statistics on the final outputs to evaluate the detection efficiency. Here, the semicolon separates the occupations in the top and bottom rows of the plaquette, and the comma separates the site occupations within each row. The obtained detection efficiencies for different spin states in different spatial locations are listed in TAB.~\ref{table:errorbudget}.

\section{Atomic spin superexchange and the two-atom entangled Bell state}

\begin{figure}[htb]
    \centering     %
    \includegraphics[width=80mm]{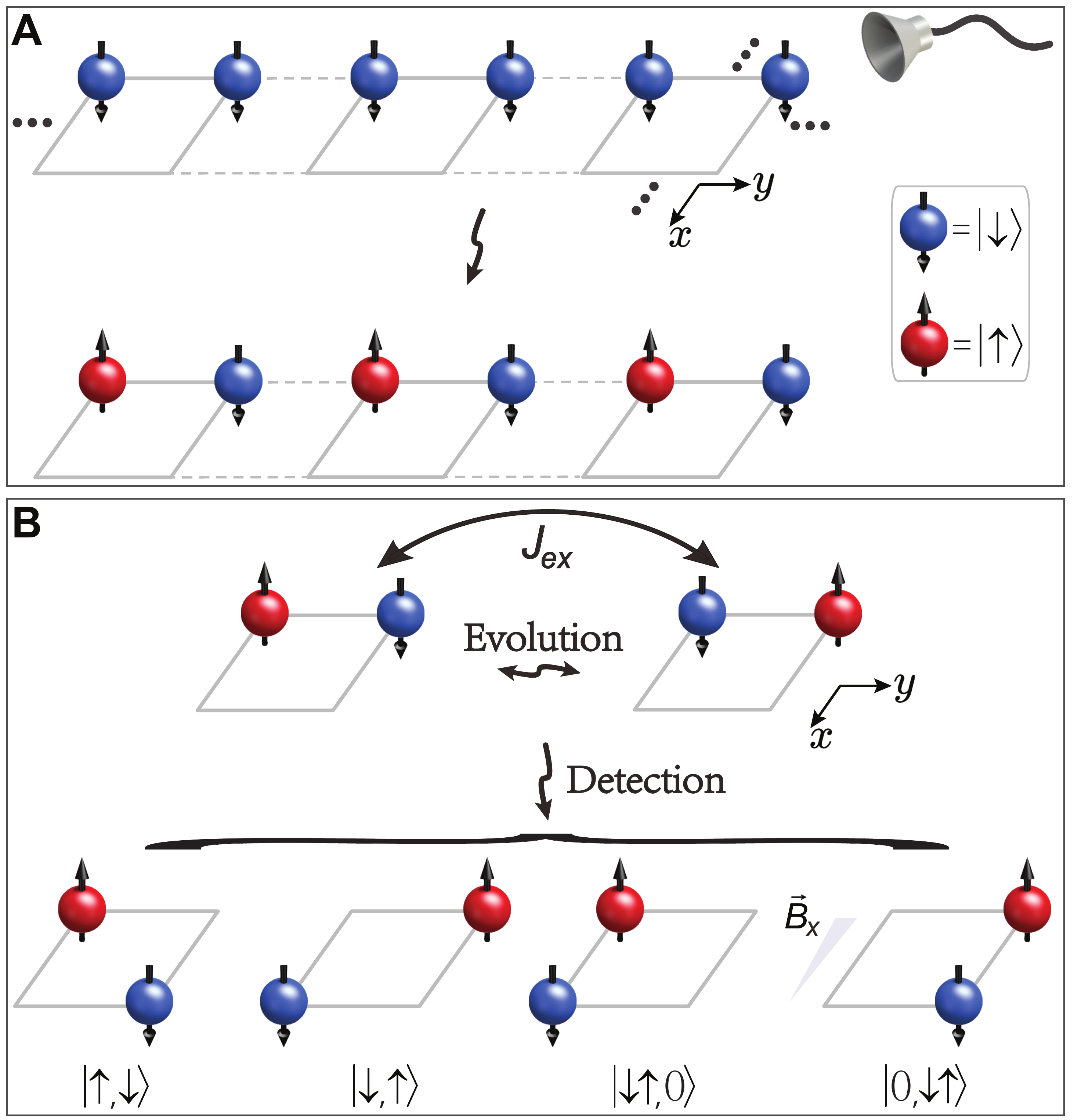}
    \caption{\textbf{State preparation and detection in atomic spin superexchange process}. (\textbf{A}) Cartoon sketch of the initial spin state preparation to study spin superexchange dynamics in each double-well along the $y$-direction. (\textbf{B}) The final spin configuration detection uses a Stern-Gerlach-type approach via applying a magnetic field gradient along the $x$-direction.
    }
    \label{figureS7:superExch}
\end{figure}

\vspace{5pt}
\noindent\textbf{Model description}.
The following two-site Bose-Hubbard Hamiltonian can describe a pair of bosonic $^{87} \mathrm{Rb}$ atoms within two different spin states (donated by $\ket{\downarrow} = \ket{F=1,m_{\mathrm{F}}=-1}$ and $\ket{\uparrow} = \ket{F=2,m_{\mathrm{F}}=-2}$) in an isolated double-well system
\begin{equation}
\begin{aligned}
\hat{H} = &\sum_{\sigma=\uparrow, \downarrow}\left[-J\left(\hat{a}_{ \mathrm{L}\sigma}^{\dagger} \hat{a}_{ \mathrm{R}\sigma}+\hat{a}_{ \mathrm{L}\sigma} \hat{a}_{ \mathrm{R}\sigma}^{\dagger} \right)\right. \\
&\left.-\frac{1}{2} \Delta\left(\hat{n}_{ \mathrm{L}\sigma}-\hat{n}_{ \mathrm{R}\sigma}\right)\right]+U\left(\hat{n}_{\uparrow \mathrm{L}} \hat{n}_{\downarrow \mathrm{L}}+\hat{n}_{\uparrow \mathrm{R}} \hat{n}_{\downarrow \mathrm{R}}\right)
\end{aligned}
\end{equation}
\noindent where $\hat{a}_{\mathrm{L},\mathrm{R}\sigma}^{\dagger}$, $\hat{a}_{\mathrm{L},\mathrm{R}\sigma}$ are creation and annihilation operators for a bosonic atom with spin $\sigma$ in the ground state of the left and right well, respectively, $\hat{n}_{\mathrm{L},\mathrm{R}\sigma} \! = \! \hat{a}_{\mathrm{L},\mathrm{R}\sigma}^{\dagger}\hat{a}_{\mathrm{L},\mathrm{R}\sigma}$, $J$ is the tunneling strength, $\Delta$ denotes the bias potential between the wells, and $U$ is the on-site interaction energy between two spin states. In this Hamiltonian, the state of the two atoms can be represented by the superposition of the Fock states $\{ \ket{\uparrow,\downarrow}, \ket{\downarrow,\uparrow}, \ket{\uparrow \downarrow,0}, \ket{0,\uparrow \downarrow} \}$.

\begin{figure}[htb]
    \centering     %
    \includegraphics[width=85mm]{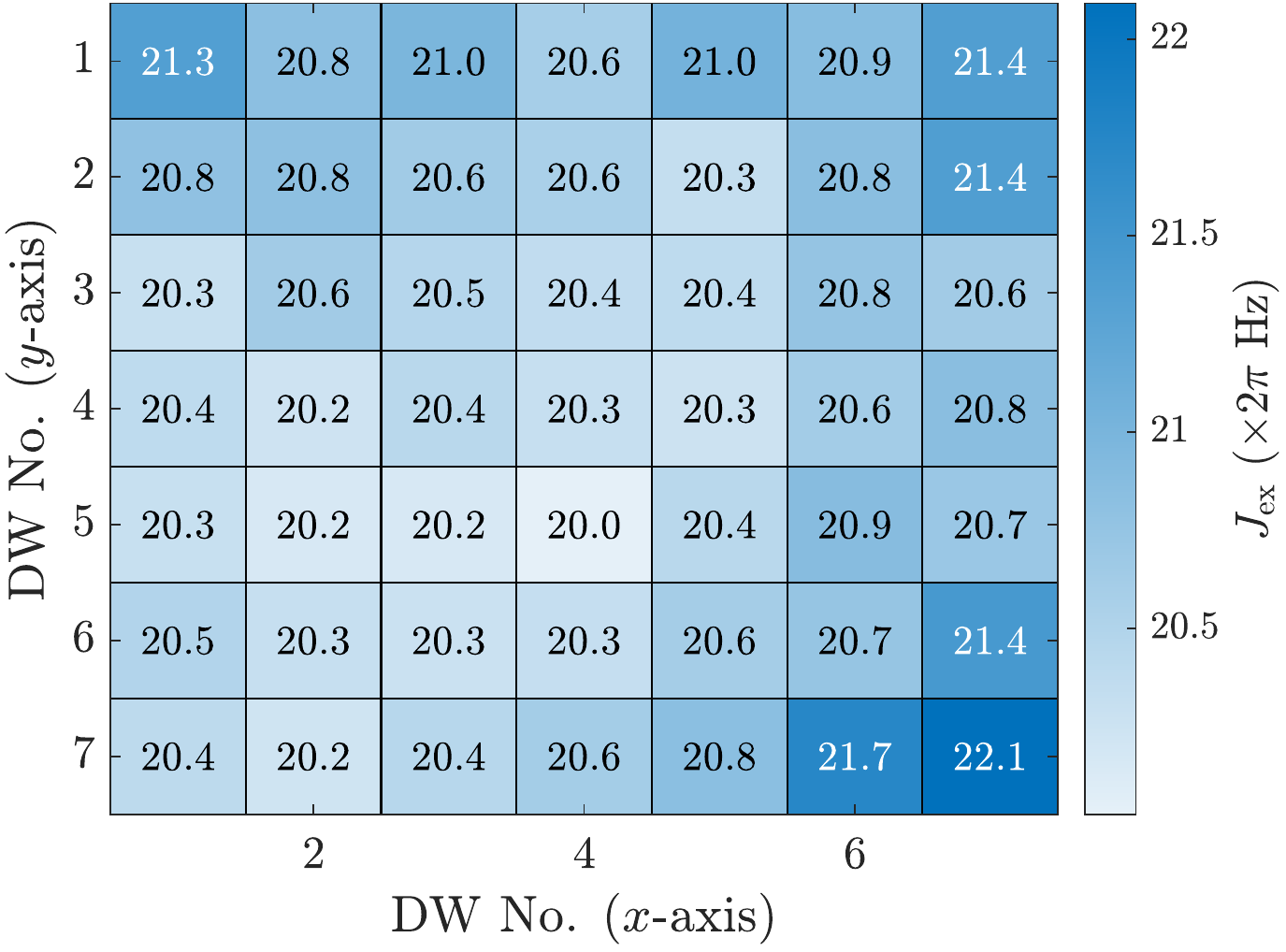}
    \caption{\textbf{}
    Map of atomic spin superexchange strength extracted from the 49 double-wells within the ROI.
    }
    \label{figureS8:JexMap}
\end{figure}

\vspace{5pt}
\noindent\textbf{Initial state preparation}.
The experiment starts by preparing a near-unity filled Mott insulator state in the staggered system. We turn on the state-dependent superlattice along the $y$-direction in $30~\mathrm{ms}$, followed by a Landau-Zener (LZ) sweep to transfer atoms in left (odd) sites from the initial state $\ket{\downarrow}$ to $\ket{\uparrow}$ described above, which creates the initial N\'eel order state $\ket{\uparrow \downarrow \uparrow \downarrow \uparrow \downarrow \cdots}$ along the $y$-direction, as shown in Fig.~\ref{figureS7:superExch}A. The initial spin configuration in each superlattice DWs is $\ket{\uparrow,\downarrow}$. The spin superexchange dynamics are started via rapidly reducing the depth of the ``short lattice'' along the $y$-direction to $18~E_{\mathrm{r}}$ in $300~\mu \mathrm{s}$, thus establishing the superexchange couplings. After a certain evolution time $t$, we freeze the spin configuration by ramping up the depth of the ``short lattice'' along $y$-direction back to $60~ E_{\mathrm{r}}$ in $300~\mu \mathrm{s}$. During the evolution dynamics, we employ a repulsive potential to compensate for the inhomogeneity of the overall trapping potential.

\vspace{5pt}
\noindent\textbf{Final state detection}.
After the atomic spin superexchange dynamics, the final spin configuration is read out using the spin-resolved detection approach mentioned above. Then, we can implement the post-selection rules that total atom number conservation and total spin conservation must be satisfied simultaneously in each plaquette to the data in analyzing atomic spin superexchange dynamics. In other words, we keep only those states in each plaquette that are in one of the four Fock states as follows, $\{ \ket{\uparrow,\downarrow}, \ket{\downarrow,\uparrow}, \ket{\uparrow \downarrow,0}, \ket{0,\uparrow \downarrow} \}$, as shown in Fig.~\ref{figureS7:superExch}B.

\noindent\textbf{Theory of Bell state fidelity}. The fidelity $\mathcal{F}$ of the generated two-atom entangled Bell pairs, $\ket{\psi} = (\ket{\uparrow,\downarrow} + \ket{\downarrow,\uparrow} ) /\sqrt{2} $, can be expressed as
\begin{equation}
\mathcal{F} = \bra{\psi} \rho \ket{\psi} = \frac{1}{4} \left(1 + \braket{\sigma_x \sigma_x} + \braket{\sigma_y \sigma_y} - \braket{\sigma_z \sigma_z} \right) 
\label{eq:fidelity2}
\end{equation}
\noindent where $\rho$ is the experimentally produced density matrix, $\sigma_{x,y,z}$ are Pauli spin operators. The bases for the $\sigma_z$, $\sigma_x$ and $\sigma_y$ operators are $\ket{\uparrow / \downarrow}$, $\ket{+/-} = (\ket{\uparrow} \pm \ket{\downarrow}) / \sqrt{2}$ and $\ket{\circlearrowleft / \circlearrowright} = (\ket{\uparrow} \pm i \ket{\downarrow}) / \sqrt{2}$, respectively. Thus, we can further deduce that
\begin{equation}
\begin{aligned}
& \braket{\sigma_x \sigma_x} = P_{+,+} + P_{-,-} - P_{+,-} - P_{-,+}, \\
& \braket{\sigma_y \sigma_y} = P_{\circlearrowleft,\circlearrowleft} + P_{\circlearrowright,\circlearrowright} - P_{\circlearrowleft,\circlearrowright} - P_{\circlearrowright,\circlearrowleft}, \\
& \braket{\sigma_z \sigma_z} = P_{\uparrow,\uparrow} + P_{\downarrow,\downarrow} - P_{\uparrow,\downarrow} - P_{\downarrow,\uparrow}.
\end{aligned}
\end{equation}
Since we have single-site- and spin-resolved detection capabilities, we can directly obtain all these two-atom spin correlations, and further derive the fidelity of Bell pairs.

\vspace{5pt}
\noindent\textbf{SPAM correction}.
Fig.~\ref{figure2:xSE}C shows the experimentally measured state populations averaged within the ROI under $\ket{+/-}$, $\ket{\circlearrowleft / \circlearrowright}$ and $\ket{\uparrow / \downarrow}$ basis without SPAM correction. The corresponding expectation values of the spin correlations are $\braket{\sigma_x \sigma_x} = 0.927 \pm 0.010$, $\braket{\sigma_y \sigma_y} = 0.924 \pm 0.014$ and $\braket{\sigma_z \sigma_z} = 0.972 \pm 0.013$, respectively. From the above, we directly deduce that the average fidelity of the generated two-atom Bell pairs is $\mathcal{F} = 0.956(5)$  over the $14 \times 14$ lattice sites region. Each experimental repetition to prepare and verify the Bell states involves an LZ-type microwave pulse to flip the spin state of those atoms in the superfluid reservoirs after the staggered-immersion cooling, followed by a resonant laser pulse remove them. Then another LZ-type microwave pulse is employed to prepare the initial N\'eel order state, and finally, applying a Stern-Gerlach-type approach to detect the spin configurations. Taking into account the SPAM correction, as shown in TAB.~\ref{table:errorbudget}, we evaluate the average fidelity of the 49 entangled Bell pairs to be $\mathcal{F} = 0.966(5)$.

\begin{figure*}[htb]
    \centering     %
    \includegraphics[width=175mm]{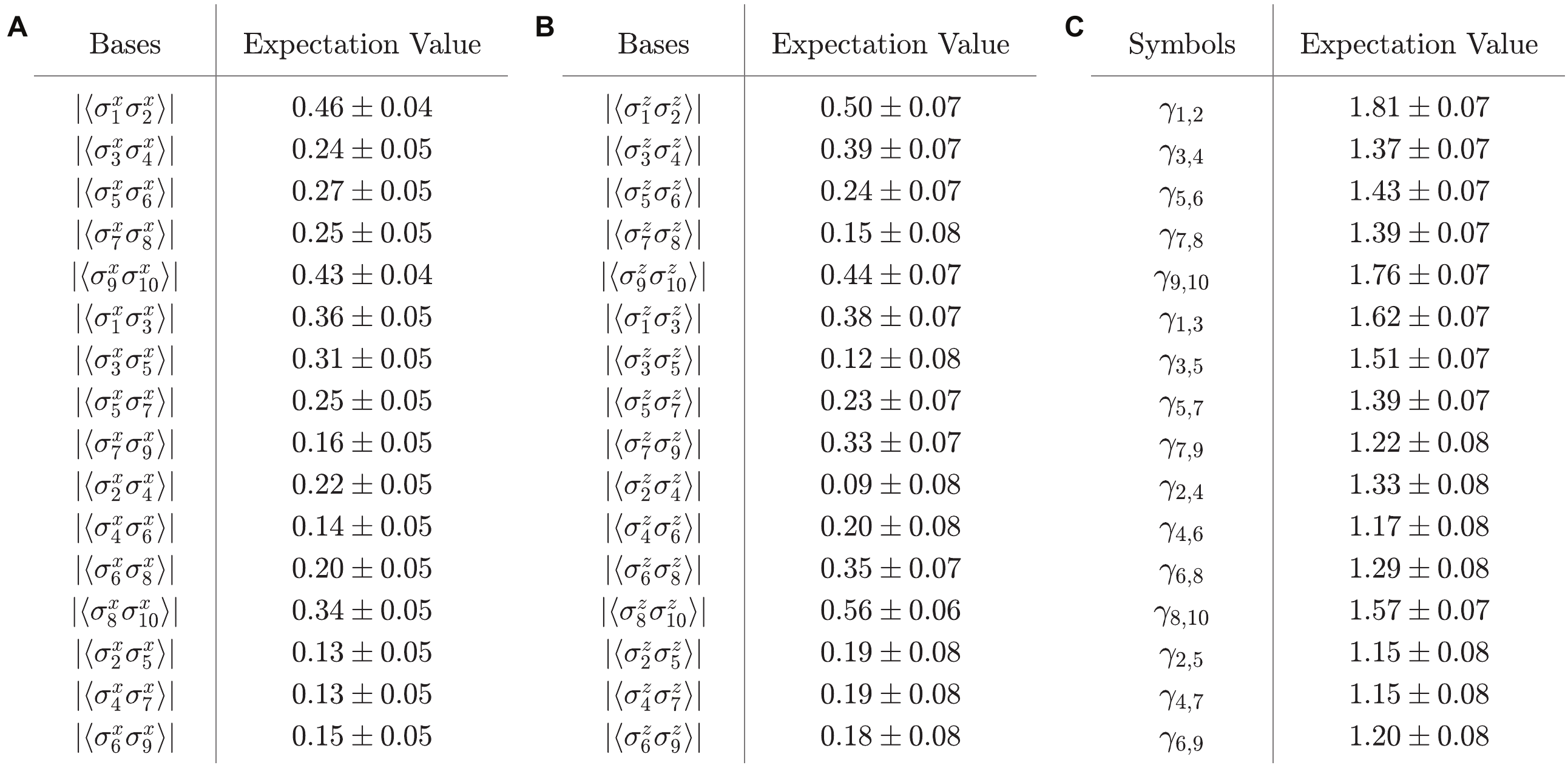}
    \caption{\textbf{Two-body spin correlations}. 
    (\textbf{A}) Configurations of two-body spin correlations under $\sigma^x$ basis, and the corresponding experimentally measured expectation values. (\textbf{B}) Configurations of two-body spin correlations under $\sigma^z$ basis, and the corresponding experimentally measured expectation values. (\textbf{C}) The expectation values of $\gamma_{i,j} =  2 \left| \left\langle \sigma_x^{i} \sigma_x^{j} \right\rangle \right| + \left| \left\langle \otimes_{k=1}^{10} \sigma_z^{k} \right\rangle \right|$ extracted from (A), where the experimentally measured ten-body spin correlation $\left| \left\langle \otimes_{k=1}^{10} \sigma_z^{k} \right\rangle \right| = 0.89 \pm 0.03$. Error bars represent the s.e.m.
    }
    \label{figureS9:spinCorr2Body}
\end{figure*}

\section{Witness and measurement settings of the multipartite entanglement}

\subsection{Witness and proof}
In previous work \cite{Zhou:2022qg}, we proposed a practical scheme for generating and detecting 1D global entanglement in the optical lattice. However, verifying the genuine multipartite entanglement (GME) property of the generated states requires rather complicated measurement settings. Therefore, in this paper, we provide a novel witness criterion for efficiently verifying the full entanglement properties of our experimentally generated 1D chains as well as the 2D plaquette state.

\vspace{5pt}
\noindent\textbf{Fully entangled state}.
According to Ref.~\cite{wang:201816,Zhou:2022qg}, an $N$-qubit quantum state $\rho$ is \textit{fully entangled}, if it cannot be expressed as following 
\begin{equation}
\rho \neq \sum_{i} p_{i} \rho_{M}^{i} \otimes \rho_{\bar{M}}^{i}
\end{equation}
\noindent for all possible bipartitions $M \vert \bar{M} $ of the whole system.

Based on our previous work \cite{Zhou:2022qg}, we developed the following witness to detect the full entanglement property, which is more experiment-friendly.

\noindent \textbf{Theorem} \textit{For an $N$-qubit quantum state $\rho$ with $N$ being an even number, let $A$ be its subsystem with an even number of particles, and let $B_x$, $B_y$ and $B_z$ be the three subsystems of $\bar{A}$. If the quantum state can be bi-separable into $M \vert \bar{M} $, where $M$ and $A$ contain an odd number of shared particles. The following inequality holds,}
\begin{equation}
| \langle \sigma_x^{A} \otimes \sigma_z^{B_x} \rangle | + | \langle \sigma_y^{A} \otimes \sigma_z^{B_y} \rangle | + | \langle \sigma_z^{A \cup B_z} \rangle | \le 1,
\end{equation}
\noindent \textit{where $\braket{O} = \mathrm{Tr}(\rho O)$ is the expectation value.}

\noindent \textbf{Proof} The proof of this theorem is similar to our previous work \cite{Zhou:2022qg}, and the details are as follows.

First, if $\rho$ is a pure quantum state under the bipartition $M \vert \bar{M} $, with $\ket{\phi} = \ket{\phi_M} \ket{\phi_{\bar{M}}}$, then 
\begin{widetext}  
\begin{eqnarray}
{} & {} & | \langle \sigma_x^{A} \otimes \sigma_z^{B_x} \rangle | + | \langle \sigma_y^{A} \otimes \sigma_z^{B_y} \rangle | + | \langle \sigma_z^{A \cup B_z} \rangle | \nonumber\\
= & {} & | \langle \sigma_x^{A \cap M} \otimes \sigma_z^{B_x \cap M} \rangle | | \langle \sigma_x^{A \cap \bar{M}} \otimes \sigma_z^{B_x \cap \bar{M}} \rangle | +  | \langle \sigma_y^{A \cap M} \otimes \sigma_z^{B_y \cap M} \rangle | | \langle \sigma_y^{A \cap \bar{M}} \otimes \sigma_z^{B_y \cap \bar{M}} \rangle | + | \langle \sigma_z^{(A \cup B_z) \cap M} \rangle | | \langle \sigma_z^{(A \cup B_z) \cap \bar{M}} \rangle | \nonumber\\
\le & {} & \sqrt{| \langle \sigma_x^{A \cap M} \otimes \sigma_z^{B_x \cap M} \rangle |^2 + | \langle \sigma_y^{A \cap M} \otimes \sigma_z^{B_y \cap M} \rangle |^2 + | \langle \sigma_z^{(A \cup B_z) \cap M} \rangle |^2} \times \nonumber \\
{} & {} & \sqrt{| \langle \sigma_x^{A \cap \bar{M}} \otimes \sigma_z^{B_x \cap \bar{M}} \rangle |^2 + | \langle \sigma_y^{A \cap \bar{M}} \otimes \sigma_z^{B_y \cap \bar{M}} \rangle |^2 + | \langle \sigma_z^{(A \cup B_z) \cap \bar{M}} \rangle |^2},
\end{eqnarray}
\end{widetext} 

\noindent where the second inequality is due to Cauchy-Schwarz inequality. Since $M$ and $A$ contain an odd number of shared particles and $A$ contians an even number of particles, then $\bar{M}$ and $A$ also contain an odd number of shared particles. Thus, one can check that $ \sigma_x^{A \cap M} \otimes \sigma_z^{B_x \cap M} $, $ \sigma_y^{A \cap M} \otimes \sigma_z^{B_y \cap M} $ and $ \sigma_z^{(A \cup B_z) \cap M} $ anticommute with each other. And $ \sigma_x^{A \cap \bar{M}} \otimes \sigma_z^{B_x \cap \bar{M}} $, $ \sigma_y^{A \cap \bar{M}} \otimes \sigma_z^{B_y \cap \bar{M}} $ and $ \sigma_z^{(A \cup B_z) \cap \bar{M}} $ anticommute with each other, too. Based on anticommutativity theorem \cite{Toth:2005ed,Asadian:2016hw}, we obtain that $| \langle \sigma_x^{A} \otimes \sigma_z^{B_x} \rangle | + | \langle \sigma_y^{A} \otimes \sigma_z^{B_y} \rangle | + | \langle \sigma_z^{A \cup B_z} \rangle | \le 1$.

On the other hand, if $\rho$ is a mixed state under such bipartition $M \vert \bar{M} $, with $\rho = \sum_k p_k \ket{\phi_k} \bra{\phi_k}$, where $\ket{\phi_k} = \ket{\phi_{Mk}} \ket{\phi_{\bar{M}k}}$ and $\sum_k p_k = 1$, then
\begin{widetext} 
\begin{eqnarray}
{} & {} & | \langle \sigma_x^{A} \otimes \sigma_z^{B_x} \rangle | + | \langle \sigma_y^{A} \otimes \sigma_z^{B_y} \rangle | + | \langle \sigma_z^{A \cup B_z} \rangle | \nonumber\\
= & {} & | \mathrm{Tr} ( \rho \sigma_x^{A} \otimes \sigma_z^{B_x} )| + | \mathrm{Tr} ( \rho \sigma_y^{A} \otimes \sigma_z^{B_y}) | + | \mathrm{Tr} (  \rho \sigma_z^{A \cup B_z} ) | \nonumber \\
\le & {} & \sum_k p_k \{ | \bra{\phi_k} \sigma_x^{A} \otimes \sigma_z^{B_x} \ket{\phi_k} | + | \bra{\phi_k} \sigma_y^{A} \otimes \sigma_z^{B_y} \ket{\phi_k} | + | \bra{\phi_k} \sigma_z^{A \cup B_z} \ket{\phi_k} |\} \nonumber \\
\le & {} & \sum_k p_k = 1.
\end{eqnarray}
\end{widetext} 
\noindent As a result, we finish the proof.

\vspace{5pt}
In addition, in our previous work \cite{Zhou:2022qg}, we have demonstrated that $\sigma_x^{\otimes N}$ and $\sigma_y^{\otimes N}$ are symmetric in our target state generation method, therefore we only need to measure one of them in our experiments.

\vspace{5pt}
\noindent \textbf{Examples}
1) A 1D entangled chain containing six particles is denoted as $\{1,2,3,4,5,6\}$. Now consider one of its subsystems $A=\{1,2\}$ and take $B_x=B_y=\varnothing$, and $B_z=\bar{A}=\{3,4,5,6\}$. According to the above theorem, if the particles in A can separate into the bipartition $ M \vert \bar{M} $, then the following inequality holds,
\begin{equation}
| \langle \sigma_x^{1} \sigma_x^{2}  \rangle | + | \langle \sigma_y^{1} \sigma_y^{2} \rangle | + | \langle \otimes_{i=1}^6 \sigma_z^{i} \rangle | \le 1.
\end{equation}
In other words, if the above inequality is broken, it indicates that the particles 1 and 2 in subsystem $A$ should all be contained in $M$ or $\bar{M}$.

2) Consider the 8-particle plaquette system shown in Fig.~\ref{figure4:2DEntangledS}C of the main text. We choose one of its subsystems $A=\{1,2\}$ and take $B_x=B_y=\{ 5 \}$, and $B_z=\{3,4\}$. According to the above theorem, if the particles in A can separate into the bipartition $ M \vert \bar{M} $, then the following inequality holds,
\begin{equation}
| \langle \sigma_x^{1} \sigma_x^{2} \sigma_z^{5}  \rangle | + | \langle \sigma_y^{1} \sigma_y^{2} \sigma_z^{5} \rangle | + | \langle  \sigma_z^{1} \sigma_z^{2} \sigma_z^{3} \sigma_z^{4} \rangle | \le 1.
\end{equation}
Similarly, if the above inequality is broken, it indicates that the particles 1 and 2 in subsystem $A$ should all be contained in $M$ or $\bar{M}$.

\vspace{5pt}
\noindent \textbf{GME witness for four-body state.}
As for the 2D four-body entangled state, we derive an efficient witness to detect GME, which can be written as follows.
\begin{align}
W_4 = & 4-(\sigma_z^{1} \sigma_x^{3} \sigma_x^{4} + \sigma_z^{1} \sigma_y^{3} \sigma_y^{4} - \sigma_z^{2} \sigma_x^{3} \sigma_x^{4} - \sigma_z^{2} \sigma_y^{3} \sigma_y^{4} \notag \\
& - \sigma_x^{1} \sigma_x^{2} \sigma_z^{3} - \sigma_y^{1} \sigma_y^{2} \sigma_z^{3} + \sigma_x^{1} \sigma_x^{2} \sigma_z^{4} + \sigma_y^{1} \sigma_y^{2} \sigma_z^{4}). 
\end{align}

\vspace{5pt}
\noindent \textbf{Proof} Now we prove abovementioned GME witness by showing that its expectation value is nonnegative for any bi-separable pure state. We begin with bipartition $1|234$. For a pure state $\ket{\phi}=\ket{\phi_1}\ket{\phi_{234}}$, we have
\begin{align}
    \langle W_4 \rangle = & 4 - (\langle \sigma_z^{1} \rangle \langle \sigma_x^{3} \sigma_x^{4} \rangle + \langle \sigma_z^{1} \rangle \langle \sigma_y^{3} \sigma_y^{4} \rangle - \langle \sigma_z^{2} \sigma_x^{3} \sigma_x^{4}  \rangle \notag \\
    & - \langle \sigma_z^{2} \sigma_y^{3} \sigma_y^{4} \rangle - \langle \sigma_x^{1} \rangle \langle \sigma_x^{2} \sigma_z^{3} \rangle - \langle \sigma_y^{1} \rangle \langle \sigma_y^{2} \sigma_z^{3} \rangle \notag \\
    & + \langle \sigma_x^{1} \rangle \langle \sigma_x^{2} \sigma_z^{4} \rangle + \langle \sigma_y^{1} \rangle \langle \sigma_y^{2} \sigma_z^{4} \rangle).
\end{align}
We denote $x, y, z$ as $\langle \sigma_x^{1} \rangle, \langle \sigma_y^{1} \rangle, \langle \sigma_z^{1} \rangle$ and define
\begin{align}
F(x,y,z) = & 4 + x (\sigma_x^{2} \sigma_z^{3} - \sigma_x^{2} \sigma_z^{4}) + y (\sigma_y^{2} \sigma_z^{3} - \sigma_y^{2} \sigma_z^{4})\notag \\
& - z (\sigma_x^{3} \sigma_x^{4} + \sigma_y^{3} \sigma_y^{4}) + \sigma_z^{2} \sigma_x^{3} \sigma_x^{4} + \sigma_z^{2} \sigma_y^{3} \sigma_y^{4}.
\end{align}
Then $\langle W_4 \rangle$ is bounded from the minimum eigenvalue of $F(x,y,z)$ \cite{Toth:2005ed}:
\begin{equation}
\langle W_4 \rangle \ge 2 - 2 \sqrt{x^2+y^2+z^2} \ge 0.
\end{equation}
The bipartitions $2|134$, $3|124$, $4|123$ can be proved similarly.

For bipartitions $12|34$, $13|24$ and $14|23$, it can be proved that $\langle W_4 \rangle \ge 0$ based on anticommutativity theorem directly.
As a result, we finish the proof.

Considering the symmetry, we only need to verify
\begin{align}
    \label{equation:S11}
    \langle \sigma_z^{1} \sigma_x^{3} \sigma_x^{4} \rangle - \langle \sigma_z^{2} \sigma_x^{3} \sigma_x^{4}  \rangle - \langle \sigma_x^{1} \sigma_x^{2} \sigma_z^{3} \rangle + \langle \sigma_x^{1} \sigma_x^{2} \sigma_z^{4} \rangle > 2
\end{align}
in experiment to demonstrate the four-body genuine entanglement.

\vspace{5pt}
\noindent \textbf{Detecting entanglement between two chains.}
If a system is separable with respect to the bipartition $1234|5678$, the following inequality holds,
\begin{equation}
| \langle \sigma_z^{3} \sigma_x^{5} \sigma_x^{7}  \rangle | + | \langle \sigma_x^{1} \sigma_x^{3} \sigma_z^{7} \rangle | + | \langle  \sigma_x^{1} \sigma_y^{3} \sigma_x^{5} \sigma_y^{7} \rangle | \le 1,
\end{equation}
which can be also proved based on anticommutativity theorem. That means we can certify entanglement between two chains by verifying $O_2 > 1$ in Fig.~\ref{figure4:2DEntangledS}E. Considering the symmetry, $O_1 > 1$ also indicates the non-separability between such two chains.

\subsection{Measurement settings and implementation}

\vspace{5pt}
\noindent \textbf{(1) $\sigma_x^1 \sigma_x^2 \sigma_z^3 \sigma_z^4$ measurement.}
To generate the entangled plaquette, we start by preparing the initial spin state in configuration $\ket{\uparrow,\downarrow;\downarrow,\uparrow}$ (the semicolon separates the occupations in the top and bottom rows of the plaquette, and the comma separates the site occupations within each row), which is achieved by our cooling approach combined with versatile potential tailoring and spin manipulation capabilities. The remaining operations are detailed in the main text.

We then present the basic procedures for implementing the $\sigma_x^1 \sigma_x^2 \sigma_z^3 \sigma_z^4$ measurement in our experiment, as illustrated in Fig.~\ref{figure4:2DEntangledS}B. After preparing such a four-body entangled state in the isolated plaquettes, we first turn on the spin-dependent effect along the $x$-direction to induce an energy difference between the left and the right sites in each double well. Then, we apply a $\pi/2$ rotation pulse to the two top sites in the plaquettes (the labeled sites 1 and 2). After that, we conduct the 2D spin-resolved detection aforementioned to capture the corresponding spin states. Similarly, the only modification in implementing the $\sigma_z^1 \sigma_z^2 \sigma_x^3 \sigma_x^4$ measurement is that the $\pi/2$ pulse is applied to the two bottom sites of the plaquette rather than the two tops.

\vspace{5pt}
\noindent \textbf{(2) $\langle \sigma_x^1 \sigma_y^3 \sigma_x^5 \sigma_y^7 \rangle $ correlation.} To generate the 2D eight-body entangled plaquette state, we start by preparing the initial spin state in configuration $\ket{\uparrow,\downarrow,\uparrow,\downarrow;\downarrow,\uparrow,\downarrow,\uparrow}$. The remaining operations are also detailed in the main text. The procedures for implementing the $\sigma_x^1 \sigma_x^2 \sigma_x^3 \sigma_x^4 \sigma_z^5 \sigma_z^6 \sigma_z^7 \sigma_z^8$ and $\sigma_z^1 \sigma_z^2 \sigma_z^3 \sigma_z^4 \sigma_x^5 \sigma_x^6 \sigma_x^7 \sigma_x^8$ measurements are the same as above. The $\langle \sigma_x^1 \sigma_y^3 \sigma_x^5 \sigma_y^7 \rangle $ correlation is extracted from the remaining measurement setting after the LOCC operation (as mentioned in the main text). The corresponding procedures are detailed as follows.

After preparing such a 2D eight-body entangled plaquette state, as illustrated in Fig.~\ref{figure4:2DEntangledS}C, following a LOCC operation, we utilize the spin-dependent superlattice potential along the $y$-direction to accumulate a $\pi/2$ phase between the adjacent atom pairs. Because of the periodic property of the optical superlattices, lattice sites labeled as 1 and 3, for example, are spatially equivalent. Therefore, no phase difference accumulates between these two lattice sites at a spatial distance of 2$a_{\mathrm{s}}$ in the presence of a spin-dependent effect produced by the superlattice structure.

However, in the 2D eight-body plaquette system illustrated in Fig.~\ref{figure4:2DEntangledS}C, when the third parallel layers of entangling gates detailed in the main text are applied, lattice sites 2 and 3 are in the identical isolated double well, as are lattice sites 6 and 7. As a result, only one atom occupies each of these lattice sites. And at this moment, site 1 and its nearest left-neighbor site share an identical double well, as does site 5. Similarly, site 4 and its nearest right-neighbor site share an identical double well, as does site 8.

Therefore, when applying the third parallel layer of entangling gates (or the LOCC operation), the atom original in site 1 (or 4, 5, 8) may tunnel to the adjacent unoccupied side of the identical double well. Especially when atoms in sites 1 and 5 are both tunneled to the left side before performing the final measurement setting, in this case, these two atoms will accumulate an additional $\pi/2$ phase compared to the atoms in sites 3 and 7 when evolving for a fixed time. After the relative phase accumulation process, we employ a $\pi/2$ rotation MW pulse before the final projection measurement.

\end{document}